\documentclass[twocolumn]{aastex61}
\usepackage{apjfonts}

\accepted{November 17, 2017}

\shorttitle{the Origin of Recombining Plasma in IC~443}
\shortauthors{Matsumura et al.}

\begin{document}

\title{Toward the Understanding of the Physical Origin of Recombining Plasma \\in the Supernova Remnant IC~443}

\correspondingauthor{Hideaki Matsumura}
\email{matumura@cr.scphys.kyoto-u.ac.jp}

\author{Hideaki Matsumura}
\affil{Department of Physics, Kyoto University, Kitashirakawa Oiwake-cho, Sakyo, Kyoto, Kyoto 606-8502, Japan}
	
\author{Takaaki Tanaka}
\affil{Department of Physics, Kyoto University, Kitashirakawa Oiwake-cho, Sakyo, Kyoto, Kyoto 606-8502, Japan}
	
\author{Hiroyuki Uchida}
\affil{Department of Physics, Kyoto University, Kitashirakawa Oiwake-cho, Sakyo, Kyoto, Kyoto 606-8502, Japan}	

\author{Hiromichi Okon}
\affil{Department of Physics, Kyoto University, Kitashirakawa Oiwake-cho, Sakyo, Kyoto, Kyoto 606-8502, Japan}

\author{Takeshi Go Tsuru}
\affil{Department of Physics, Kyoto University, Kitashirakawa Oiwake-cho, Sakyo, Kyoto, Kyoto 606-8502, Japan}


\begin{abstract}
We perform a spatially resolved spectroscopic analysis of X-ray emission from the supernova remnant (SNR) IC~443 with {\it Suzaku}. 
All the spectra are well reproduced by a model consisting of a collisional ionization equilibrium (CIE) and two recombining plasma (RP) components. 
Although previous X-ray studies found an RP in the northeastern region, this is the first report on RPs in the other parts of the remnant. 
The electron temperature $kT_e$ of the CIE component is almost uniform at $\sim 0.2$~keV across the remnant. 
The CIE plasma has metal abundances consistent with solar and is concentrated toward the rim of the remnant, suggesting 
that it is of shocked interstellar medium origin.  
The two RP components have different $kT_e$: one in the range of  0.16--0.28~keV and the other in the range of 0.48--0.67~keV. 
The electron temperatures of both RP components decrease toward the southeast, where the SNR shock is known to be interacting with a molecular cloud.
We also find the normalization ratio of the lower-$kT_e$ RP to higher-$kT_e$ RP components increases toward the southeast. 
Both results suggest the X-ray emitting plasma in the southeastern region is significantly cooled by some mechanism. 
One of the plausible cooling mechanisms is thermal conduction between the hot plasma and the molecular cloud. 
If the cooling proceeds faster than the recombination timescale of the plasma, the same mechanism can account for the recombining plasma as well. 
\end{abstract}


\section{Introduction} \label{sec:intro}
IC~443 (G189.1+3.0) is a Galactic supernova remnant (SNR) at a distance of 1.5 kpc \citep{Welsh2003}. 
Its age is estimated to be in the range of 3--30 kyr \citep{Petre1988,Olbert2001}.
\cite{Olbert2001} found a pulsar wind nebula (PWN) in the southeastern part of the remnant, suggesting that IC~443 is a remnant of a core-collapse supernova.
Seen in the radio and optical bands, IC~443 has two shells with different radii \citep[e.g.,][]{Lee2008}.
The remnant is categorized as a mixed-morphology SNR \citep{Rho1998} since the X-ray emission has a center-filled morphology.

In X-rays, \cite{Kawasaki2002} measured the Ly$\alpha$ to He$\alpha$ line intensity ratio of Si and S with {\it ASCA}, and found that the plasma in the northeastern part of the remnant is more ionized 
than expected in collisional ionization equilibrium (CIE).
With {\it Suzaku}, \cite{Yamaguchi2009} discovered radiative recombining continua (RRCs) of Si and S in the northeastern region, which provided unambiguous evidence for a recombining plasma (RP).
\cite{Ohnishi2014} found RRCs of Ca and Fe in almost the same region as  \cite{Yamaguchi2009} and indicated that heavier elements are also overionized.

The formation process of RPs, now observed also in other Galactic and Large Magellanic Cloud SNRs \citep[e.g.,][]{Ozawa2009,Uchida2015}, has not been 
fully understood yet and is still under debate. 
The rarefaction scenario \citep{Itoh1989} and the thermal conduction scenario \citep{Kawasaki2002} are mainly discussed in the literature. 
\cite{Itoh1989} proposed that RPs are realized by adiabatic cooling when the SNR shock breaks out of a dense circumstellar matter into a lower density interstellar medium (ISM).
\cite{Kawasaki2002} suggested that the electron temperature ($kT_e$) of the plasma is lowered by thermal conduction between the SNR plasma and a molecular cloud interacting with the SNR.
With {\it Suzaku} observations of the SNR G166.0+4.3, \cite{Matsumura2017} found an RP only in a part of the remnant with higher ambient gas density, which favors the thermal conduction scenario. 
On the other hand, with {\it Chandra} observations of W49B, \cite{Lopez2013} found that the plasma has a gradient of $kT_e$ increasing toward a region where higher density ISM is present.
They claimed that the dominant cooling mechanism is adiabatic expansion of the hot plasma.

IC~443 is interacting with both molecular and atomic gas \citep[e.g.,][]{Lee2008}.
\cite{Cornett1977} discovered a  dense molecular cloud associated with IC~443.  
CO line emissions in the southeastern part of the remnant suggest that the molecular cloud is interacting with IC~443 \citep[e.g.,][]{Denoyer1979,Xu2011,Yoshiike2017}.
Measurements of the CO line velocity by \cite{Yoshiike2017} suggest that the most part of the interacting molecular gas is located in front of IC~443. 
In the northeast, the SNR shock emits atomic lines expected from post-shock recombining gas \citep{Fesen1980}, and therefore, the shock is likely to be propagating into an atomic medium.
Near-infrared observations also support the atomic gas in the northeast \citep[e.g.,][]{Kokusho2013}.

In this paper, we perform spatially resolved X-ray spectroscopy of the SNR IC~443 with {\it Suzaku} and 
compare the results with the distribution of interstellar gas interacting with the remnant. 
We then discuss the implications particularly on the formation process of the RP. 
Throughout the paper, errors are quoted at 90\% confidence levels in the text and tables, and error bars in the figures indicate 1$\sigma$ confidence intervals.
All spectral fits are performed with XSPEC version 12.9.0n.
The plasma models are calculated with ATOMDB version 3.0.8.
We used solar abundances given by \cite{Wilms2000}.

\section{Observations and data reduction} \label{sec:obs}
Table~\ref{tab:obs_log} summarizes the {\it Suzaku} observation log of IC~443.
We used the data from the X-ray Imaging Spectrometer \citep[XIS;][]{Koyama2007}, which consists of four X-ray CCD cameras installed on the focal planes of the X-Ray Telescopes \citep[XRT;][]{Serlemitsos2007}. 
XIS0, 2 and 3 have front-illuminated sensors whereas XIS1 has a back-illuminated one.
XIS2 and a part of XIS0 have not been functioning since November 2006 and June 2009, respectively ({\it Suzaku} XIS documents\footnote{http://www.astro.isas.ac.jp/suzaku/doc/suzakumemo/suzakumemo-2007-08.pdf}$^{,}$\footnote{http://www.astro.isas.ac.jp/suzaku/doc/suzakumemo/suzakumemo-2010-01.pdf}).
Therefore, we do not use XIS2 and the part of XIS0 for the data acquired after those times.

We reduced the data using the HEADAS software version 6.19.
We used the calibration database released in April 2016 for processing the data.
We removed flickering pixels by referring to the noisy pixel maps\footnote{https://heasarc.gsfc.nasa.gov/docs/suzaku/analysis/xisnxbnew.html} provided by the XIS team.
We also discarded pixels adjacent to the flickering pixels. 
Non X-ray backgrounds (NXBs) were estimated by {\tt xisnxbgen} \citep{Tawa2008}.
Redistribution matrix files and ancillary response files were produced by {\tt xisrmfgen} and {\tt xissimarfgen} \citep{Ishisaki2007}, respectively.

Analyzing the XIS0 and 3 data taken in 2007, we noticed that the centroid energies of Ly$\alpha$ lines from IC~443, and Mn K$\alpha$ and K$\beta$ lines from the on-board calibration sources are shifted by 2--8~eV from the true values. 
Although this is well within the calibration uncertainties, $< 20$~{\rm eV} at Mn K$\alpha$, reported by the XIS team\footnote{https://heasarc.gsfc.nasa.gov/docs/suzaku/prop\_tools/suzaku\_td/}, residuals caused by the gain shift are not 
negligible with the high statistics of the present data. 
Those artificial residuals sometimes make it difficult to clearly see important residuals in the spectral fitting described in \S3. 
We, therefore, corrected the gain of the XIS0 and 3 data from each observation. 
We fitted narrow-band spectra in each energy band around the  \ion{O}{8}, \ion{Mg}{12}, \ion{Si}{14}, and \ion{S}{16} lines from IC~443, and 
Mn K$\alpha$ and K$\beta$ lines from the calibration sources with a Gaussian and a power raw for the underlying continuum. 
We here did not use the \ion{Ne}{10} line because it overlaps with the Fe-L line complex.
We then fitted the relation between the true energies and the centroid energies of the Gaussians with a first-order polynomial. 
We corrected the gain using the polynomial function.

\begin{deluxetable*}{lcccc}[ht]
\tablecaption{Observation log. \label{tab:obs_log}}
\tablecolumns{5}
\tablewidth{0pt}
\tablehead{
\colhead{Target} &
\colhead{Obs. ID} &
\colhead{Obs.~date} &
\colhead{(R.A., Dec.)} &
\colhead{Effective Exposure}
}
\startdata
IC~443 northeast & 501006010 & 2007-03-06 & ($6^{\rm{h}} 17^{\rm{m}} 11\fs4,~22\degr 46\arcmin 32\farcs$5) & 42~ks \\
IC~443 southeast & 501006020 & 2007-03-07 & ($6^{\rm{h}} 17^{\rm{m}} 11\fs3,~22\degr 28\arcmin 46\farcs$9) & 44~ks \\
IC~443 northwest & 505001010 & 2010-09-17 & ($6^{\rm{h}} 15^{\rm{m}} 59\fs4,~22\degr 45\arcmin 18\farcs$7) & 83~ks \\
IRAS 05262+4432 (Background) & 703019010 & 2008-09-14 & ($5^{\rm{h}} 29^{\rm{m}} 56\fs0,~44\degr 34\arcmin 39\farcs$2) & 82~ks \\
\enddata
\end{deluxetable*}

\section{Analysis} \label{sec:alaysis}
\subsection{Image}
\begin{figure}
\begin{center}
 \includegraphics[width=7cm]{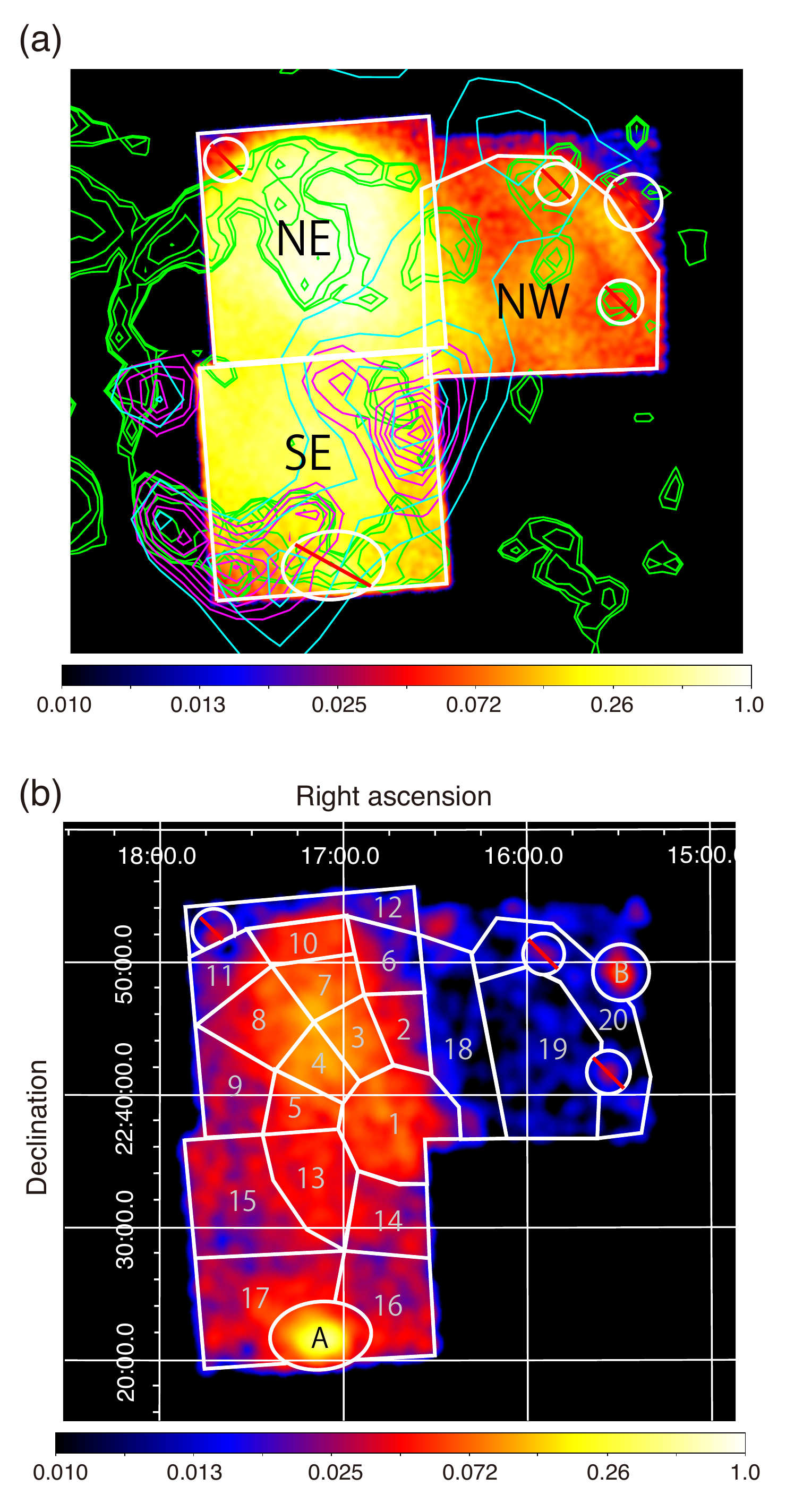} 
\end{center}
\caption{XIS3 images of IC~443 in the energy band of (a) 0.3--2.0~keV and (b) 3.0--5.0~keV after NXB subtraction and correction of the vignetting effect.
In panel (a), we overlaid green contours of 1.4~GHz radio image which were obtained from the NRAO VLA Sky Survey (NVSS).
The cyan and magenta contours are radio images of $^{12}$CO($J = 1-0$) and $^{12}$CO($J = 2-1$) obtained from the NANTEN2, respectively \citep{Yoshiike2017}.
Sources in regions A and B in panel~(b) are the PWN 1SAX J0617.1+2221 and the point source U061530.75+224910.6, respectively.
The X-ray count rates is normalized so that the peak becomes unity.
The coordinate refers to the J2000.0 epoch.
}
\label{fig:ic443_img}
\end{figure}

Figure~\ref{fig:ic443_img}a shows an XIS3 image of IC~443 in the energy band of 0.3--2.0~keV.
We subtracted the NXB from the image and corrected for the vignetting effect of the XRT.
The X-ray emissions in the east are center-filled as is previously known \citep[e.g.,][]{Troja2006}.
In the northwest, the image shows a hint of a shell-like structure.

Figure~\ref{fig:ic443_img}b shows an XIS3 image in the 3.0--5.0~keV band.
We found the PWN 1SAX J0617.1+2221 (region~A) and the point source U061530.75+224910.6 (region~B) \citep{Bocchino2001,Bocchino2003}.
We found two additional point sources in the northwest and a source in the northeast which are already detected in {\it XMM-Newton} data by \cite{Bocchino2003}. 
We excluded the PWN and the point sources for our spectral analysis of IC~443.

\subsection{X-ray Background Estimation}
We used {\it Suzaku} data of IRAS~05262+4432, which is located at (R.A.,~Dec.) $=$ ($5^{\rm{h}} 29^{\rm{m}} 56\fs0,~44\degr 34\arcmin 39\farcs$2), for background estimation.
The observation log is shown in Table~\ref{tab:obs_log}.
We extracted a spectrum from a source-free region and fitted it with a model by \cite{Masui2009}, who studied the soft X-ray emission from the anti-center region of the Galaxy with {\it Suzaku}.
The model consists of four components: the cosmic X-ray background (CXB), the local hot bubble (LHB), and two thermal components for the Galactic halo ($\rm GH_{cold}$ and $\rm GH_{hot}$).
We used the Wisconsin absorption model \citep[Wabs;][]{Morrison1983} for the absorption column density ($N_{\rm H}$) and fixed it at $3.8 \times 10^{21}~\rm cm^{-2}$, which is the total Galactic absorption in the line of sight toward IRAS~05262+4432 \citep{Dickey1990}.
We fixed the photon index of the CXB component and $kT_e$ of the LHB, $\rm GH_{cold}$ and $\rm GH_{hot}$ components at the values given by \cite{Masui2009}.
The best-fit parameters are shown in Table~\ref{tab:bkg_model}.
We used this model as the background spectrum for IC~443  in subsection~\ref{sec:alaysis_snr} but with $N_{\rm H}$ of the CXB component changed to $6.1 \times 10^{21}~\rm cm^{-2}$, which corresponds to the total Galactic absorption in the line of sight toward IC~443.

\begin{deluxetable}{cccc}[t]
\tablecaption{Best-fit model parameters for the background.
\label{tab:bkg_model}}
\tablehead{
\colhead{Component} &
\colhead{Model function} & 
\colhead{Parameter} &
\colhead{Value}
} 
\startdata
      CXB & Wabs & $N_{\rm H}$ (10$^{22}$ cm$^{-2}$) & 0.38 (fixed) \\
      & Power law & Photon index & 1.4 (fixed) \\
      & & Normalization$^{\ast}$ & 10.7 $\pm$ 0.4 \\
      LHB & APEC & $kT_e$ (keV) & 0.105 (fixed) \\
      & & Normalization$^{\dagger}$ & 13.4 $\pm$ 3.2\\
      $\rm GH_{cold}$ & APEC & $kT_e$ (keV) & 0.658 (fixed) \\
      & & Normalization$^{\dagger}$ & 2.1 $\pm$ 0.2\\
      $\rm GH_{hot}$ & APEC & $kT_e$ (keV) & 1.50 (fixed) \\
      & & Normalization$^{\dagger}$ & 3.1 $\pm$ 0.5\\
      \hline
      & & $\chi^{2}_{\nu}$ ($\nu$) & 1.29 (234) \\
\enddata
\tablecomments{
$^{\ast}$The unit is photons s$^{-1}$ cm$^{-2}$ keV$^{-1}$ sr$^{-1}$ at 1~keV.\\
$^{\dagger}$The emission measure integrated over the line of sight,\\ i.e., (1 / 4$\pi$) $\int n_e n_{\rm H} dl$ in units of 10$^{14}$ cm$^{-5}$ sr$^{-1}$.
}
\end{deluxetable}

\subsection{SNR spectra} \label{sec:alaysis_snr}
Previous X-ray studies found that the spectra in the northeastern region can be reproduced with RP models \citep{Yamaguchi2009,Ohnishi2014}.
In order to examine ionization states of the SNR plasmas in the other regions, we extracted spectra from the southeast (SE), northwest (NW) and northeast (NE) regions shown in Figure~\ref{fig:ic443_img}a.

\begin{figure}
\begin{center}
 \includegraphics[width=7.5cm]{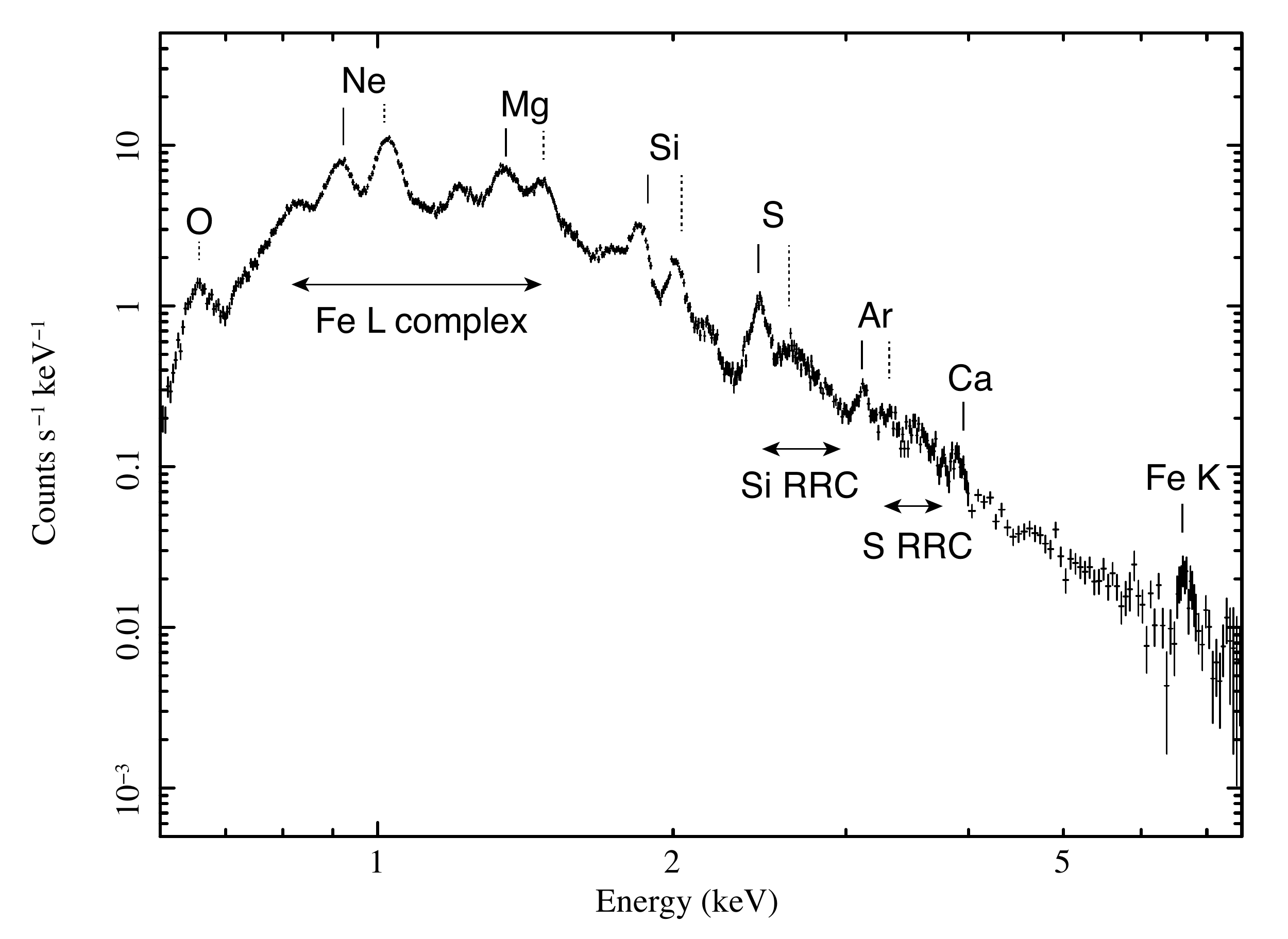} 
\end{center}
\caption{XIS0+3 spectrum of the SE region after NXB subtraction.
The vertical solid and dashed black lines indicate the centroid-energies of He$\alpha$ and Ly$\alpha$ lines, respectively.
The arrows indicate the energy bands of the Fe-L complex and the radiative recombining continua of Si and S.}
\label{fig:SE_raw}
\end{figure}

Figure~\ref{fig:SE_raw} shows the XIS0+3 spectrum of the SE region after NXB subtraction.
The characteristic structures of Si and S RRCs indicate that the plasma is in a recombination-dominant state.
Following \cite{Yamaguchi2009}, we first analyzed the spectrum in the energy band above 1.6~keV in order to quantify the ionization states of Si and S.
We applied a model consisting of one RP component to the spectrum. 
We used the VVRNEI model in XSPEC, which calculates the spectrum of a non-equilibrium ionization plasma after a rapid transition of the electron temperature from $kT_{\rm init}$ to $kT_e$.
The initial plasma temperature $kT_{\rm init}$ was fixed at 5~keV, in which most Si and S ions become bare nuclei.
The present electron temperature $kT_e$ and ionization parameter $n_et$ were allowed to vary.
The abundances of Si, S, Ar, Ca and Fe were allowed to vary, whereas the Ni abundance was linked to Fe.
The abundances of the other elements were fixed to solar. 
We used the Tuebingen-Boulder ISM absorption model \citep[TBabs;][]{Wilms2000}, whose column density was fixed at $7.0 \times 10^{21}~\rm cm^{-2}$ as determined by \cite{Kawasaki2002}.
For the XIS0+3 spectrum, we ignored the energy band of 1.78--1.92~keV because of the known calibration uncertainty around the neutral Si K-edge. 
Although we excluded the PWN from the spectrum extraction region, there is non-negligible leakage from the PWN due to the tail of the point spread function of the XRT. 
We, therefore, added a power law to account for the PWN emission. 
We fixed the photon index at 1.89 given by our spectral analysis of the PWN region (see Appendix~\ref{apx:point}).

Figure~\ref{fig:SE_spe}a and Table~\ref{tab:spe_model} show the fitting results and the best-fit parameters, respectively.
The spectrum is reproduced well by the model with $kT_e = 0.30 \pm 0.01~\rm keV$ and $n_et = (4.6 \pm 0.1) \times 10^{11}~\rm s~cm^{-3}$.
The electron temperature is significantly smaller than $0.61_{-0.02}^{+0.03}$~keV given by \cite{Yamaguchi2009}. 
They fitted the spectrum with a phenomenological model consisting of a CIE component, Gaussians for Si, S and Ar Ly$\alpha$ lines and RRCs of Mg, Si and S.
We here used a more physically oriented model that takes into account ion fractions and ionization timescales of all elements in the recombination-dominant state. 
In this model, RRCs of Ne and of lighter elements, which are not included in the model employed by \cite{Yamaguchi2009}, extend up to the Si--S band. 
This difference would be the reason why we obtained lower $kT_e$. 

We extrapolated the above model down to 0.6 keV, which results in large residuals in the soft band. 
We added a CIE component to the model following \cite{Kawasaki2002} and \cite{Bocchino2009}, who modeled IC~443 spectra with two-component models with a CIE component for the soft band. 
The result is shown in Figure~\ref{fig:SE_spe}b and Table~\ref{tab:spe_model}.
Since we found the abundances of the CIE component are close to solar, we fixed them at solar values.
The fit left residuals around 1.23 keV, most probably due to the lack of Fe-L lines in the model \citep[e.g.,][]{Yamaguchi2011}, and therefore, we added a Gaussian at 1.23 keV.
The fit still left residuals at 0.87~keV, indicating a shortage of the emission line of \ion{Fe}{18} 1s2 2s2 2p4 3d1 $\rightarrow$ 1s2 2s2 2p5.
To make the line stronger with $n_et \sim 10^{11}~\rm s~cm^{-3}$, $kT_e$ is required to be higher than $\sim 0.5$~keV.

We then fitted the spectrum with a model consisting of a power-law, a CIE, an RP and an additional higher $kT_e$ components.
We tried CIE, ionizing plasma (IP), and RP models as the high $kT_e$ component, and found that the CIE and IP models fail to fit the spectrum. 
We applied a model consisting of a CIE and two RP components to the spectrum.
The spectrum is well reproduced by the model with the significantly different $kT_e$ between the two RP components 
whereas $n_et$ and abundances of the two components are almost the same.
Therefore, we linked $n_et$ and the abundances of the components to each other.
The best-fit model and parameters are shown in Figure~\ref{fig:SE_spe}c and Table~\ref{tab:spe_model}, respectively.
The residuals at the \ion{Fe}{18} line was significantly improved.
The electron temperatures of the CIE and two RP (named $\rm RP_{cold}$ and $\rm RP_{hot}$) components are $0.22 \pm 0.01$~keV, $0.19 \pm 0.01$~keV and $0.54_{-0.01}^{+0.03}$~keV, respectively.
The ionization timescale of the RP components is $n_et = (4.2_{-0.1}^{+0.2}) \times 10^{11}~\rm s~cm^{-3}$.

We performed spectral fitting of the NW and NE spectra in the same procedure as we did for the SE spectrum. 
We found that the model consisting of a CIE and two-RP components gives satisfactory fits to both spectra. 
The spectra with the best-fit models are plotted in Figures~\ref{fig:NE_NW_spe}a and \ref{fig:NE_NW_spe}b and the best-fit parameters are again summarized in Table~\ref{tab:spe_model}. 
The ionization parameter $n_et$ of the NW spectrum is close to $2 \times 10^{12}~\rm s~cm^{-3}$, which is a characteristic timescale for a plasma to reach CIE \citep{Masai1994}.
Therefore, the RP in the NW is close to CIE.
Since we cannot fit the spectrum in the energy band of the Fe-L complex of the NW spectrum, we allowed the Fe abundance of the CIE component to vary. 
The fit gave somewhat higher Fe abundance of $1.3 \pm 0.1$, but the abundance is still close to solar. 
Although $kT_e$ and $n_et$ of the NW and NE spectra are close to those of the SE spectrum, normalization ratios of the RP components are significantly different, indicating that the RP components have spatial variation.

\begin{figure}
\begin{center}
 \includegraphics[width=7cm]{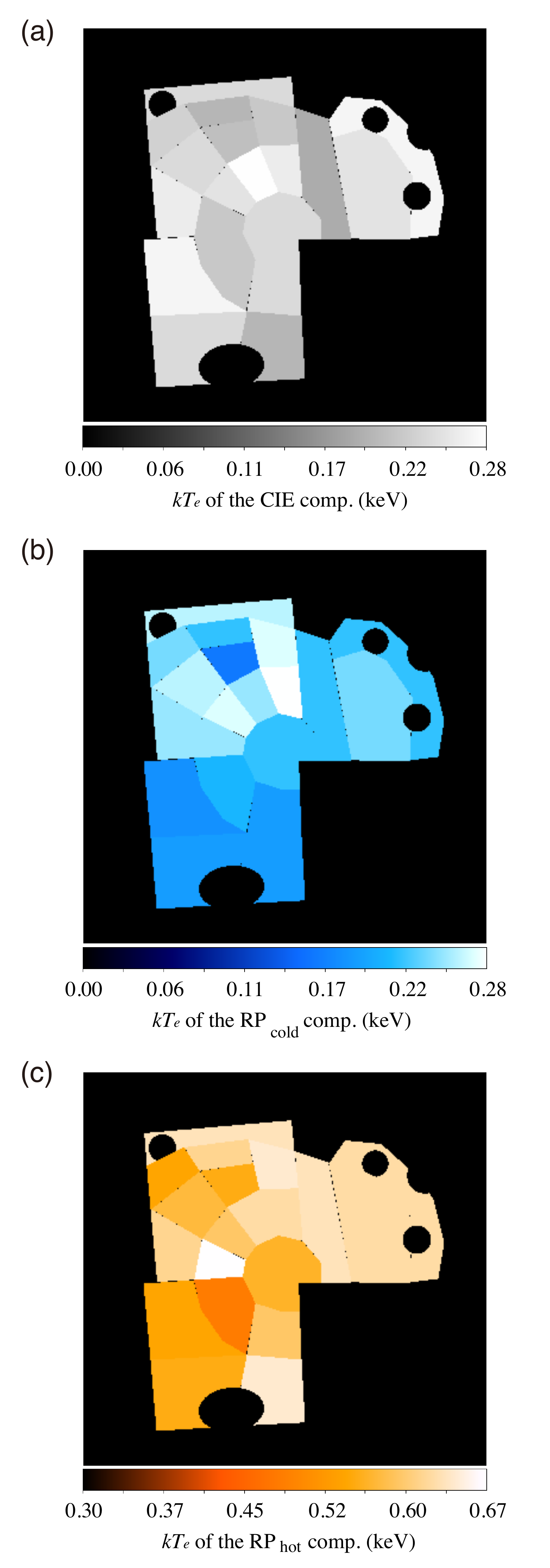} 
\end{center}
\caption{Maps of the electron temperatures $kT_e$ of the (a) CIE, (b) $\rm RP_{cold}$, (c) $\rm RP_{hot}$ components.
}
\label{fig:kTe}
\end{figure}

\begin{figure}
\begin{center}
 \includegraphics[width=6.5cm]{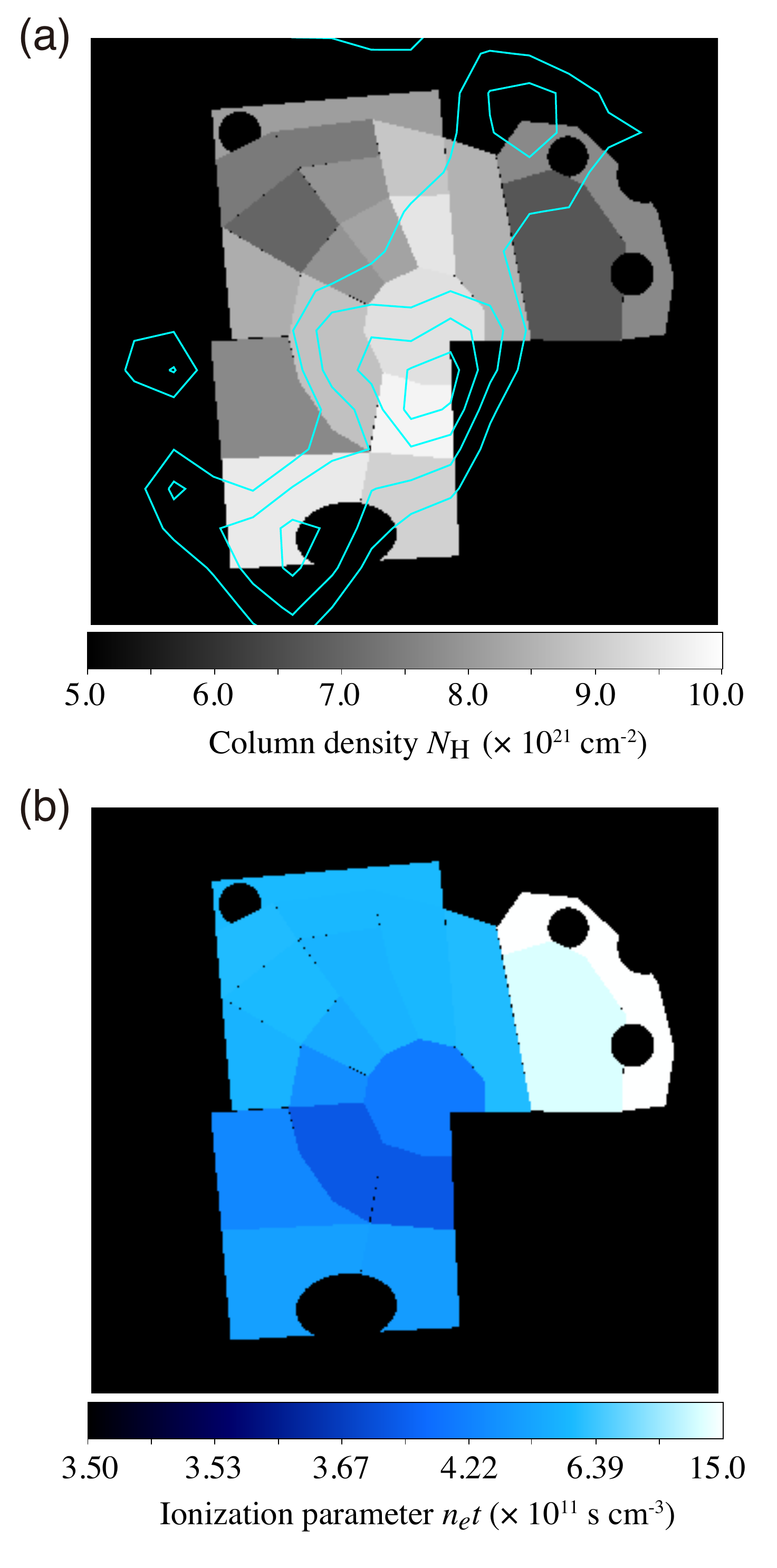} 
\end{center}
\caption{(a) Map of the column density $N_{\rm H}$ with the $^{12}$CO($J = 1-0$) emissions \citep[cyan contours;][]{Yoshiike2017}.
(b) Map of the ionization parameters $n_et$.}
\label{fig:fitting_result1}
\end{figure}

In order to conduct spatially resolved spectroscopy in greater detail, we divided the remnant into 20 regions as shown in Figure~\ref{fig:ic443_img}b and extracted spectra from each region. 
We fitted the spectra with the above model, and found that it reproduces all the spectra well. 
The details of the fits are described in Appendix~\ref{apx:1-20}. 
Since the leakage from the PWN and the point sources contaminate the spectra of regions 16, 17 and 20, we added a power law to the model for these spectra. 
The photon indices were fixed at 1.89 and 2.22, which are determined by our analysis in Appendix~\ref{apx:point}, for the PWN and the point-source U061530.75+224910.6, respectively.
The obtained ranges of $kT_e$ are 0.19--0.28~keV, 0.16--0.28~keV and 0.48--0.67~keV for the CIE, $\rm RP_{cold}$ and $\rm RP_{hot}$ components, respectively.
The electron temperature of the CIE component is uniform whereas those of the $\rm RP_{cold}$ and $\rm RP_{hot}$ components decrease toward the southeast as shown in Figure~\ref{fig:kTe}.
Figures~\ref{fig:fitting_result1}a and \ref{fig:fitting_result1}b show the map of $N_{\rm H}$ and $n_et$, respectively.
In the northwest, $n_et$ is larger than those of the other regions.
~~~\\
~~~\\

\begin{figure}
\begin{center}
 \includegraphics[width=7.5cm]{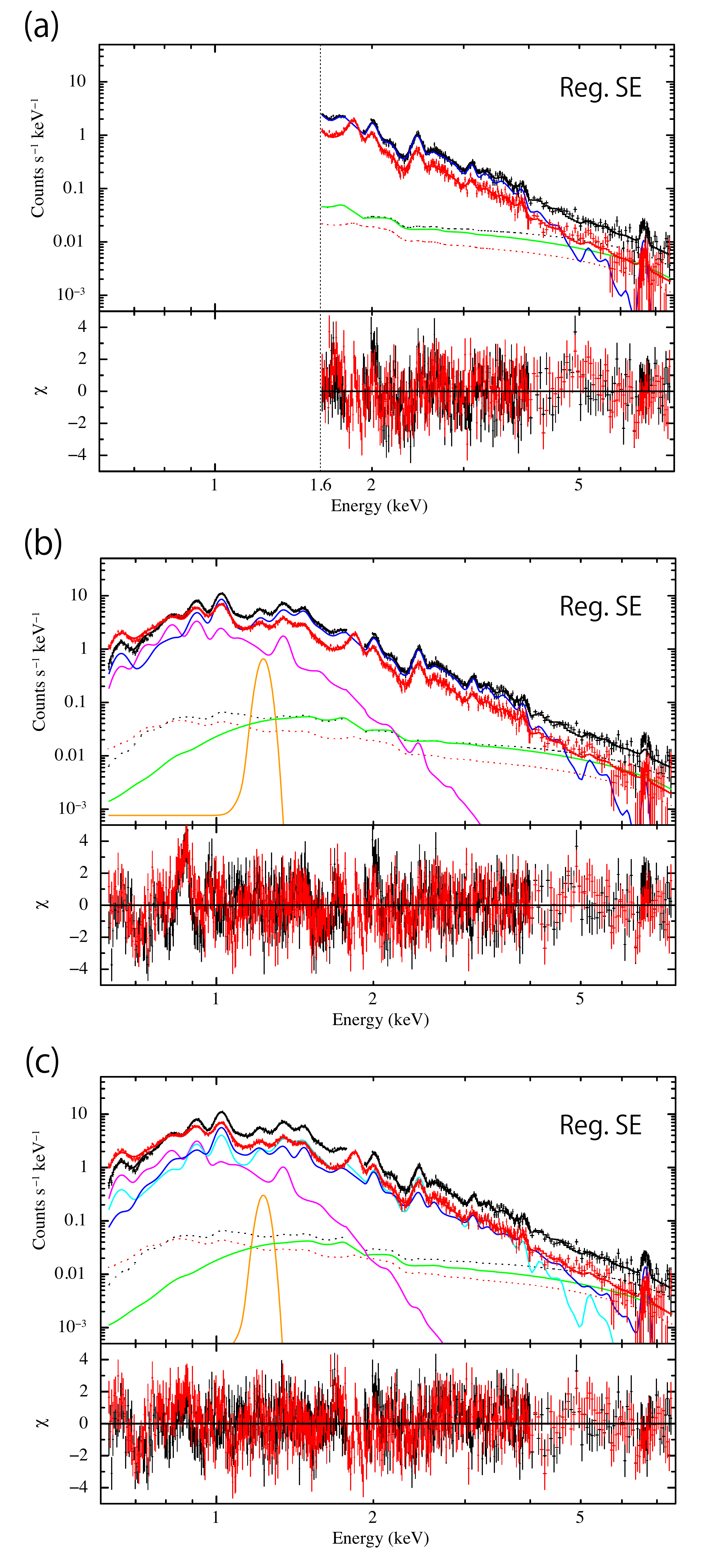} 
\end{center}
\caption{XIS0+3 (black crosses) and XIS1 (red crosses) spectra in the SE region.
Each spectrum is fitted with the model consisting of the SNR and background components.
The magenta, cyan, blue, green, and orange solid lines show the SNR models for XIS0+3, consisting of CIE, $\rm RP_{cold}$, $\rm RP_{hot}$, power-law, Gaussian components, respectively.
The black and red dotted lines denote the background model for the XIS0+3 and XIS1 spectra, respectively.
The sum of the SNR and background models for XIS0+3 and XIS1 are shown by the black and red solid lines, respectively.
}
\label{fig:SE_spe}
\end{figure}

\begin{figure}
\begin{center}
 \includegraphics[width=7.5cm]{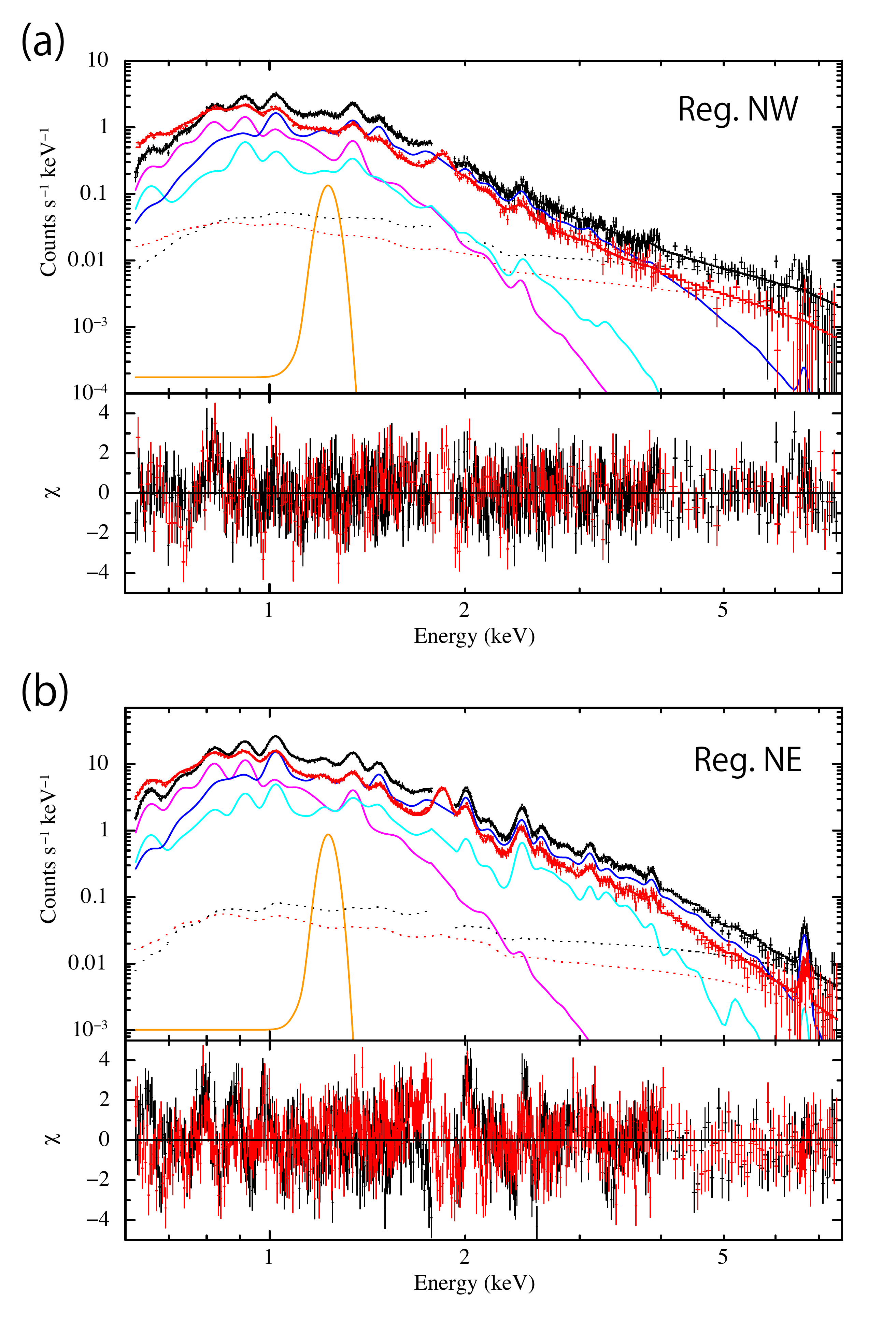} 
\end{center}
\caption{Same as Figure~\ref{fig:SE_spe} but for the (a) NW and (b) NE spectra.
}
\label{fig:NE_NW_spe}
\end{figure}

\begin{deluxetable*}{ccccccc}[t]
\tablecaption{Best-fit model parameters of the SE, NW and NE spectra.
\label{tab:spe_model}}
\tablehead{
\colhead{} & 
\colhead{} & 
\colhead{SE} &
\colhead{SE} &
\colhead{SE} &
\colhead{NW} &
\colhead{NE}\\
\colhead{Model function} & 
\colhead{Parameter} & 
\colhead{1.6--7.5 keV} &
\colhead{0.6--7.5 keV} &
\colhead{0.6--7.5 keV} &
\colhead{0.6--7.5 keV} &
\colhead{0.6--7.5 keV}
} 
\startdata
      TBabs & $N_{\rm H}$ (10$^{22}$ cm$^{-2}$) & 0.70 (fixed) & 0.90 $_{-0.01}^{+0.02}$ & 0.88 $_{-0.02}^{+0.01}$ & 0.74 $_{-0.03}^{+0.02}$ & 0.85 $_{-0.01}^{+0.01}$ \\
      VVRNEI1 & $kT_e$ (keV) & ------------ & 0.28 $_{-0.03}^{+0.01}$ & 0.22 $_{-0.01}^{+0.01}$ & 0.26 $_{-0.02}^{+0.02}$ & 0.23 $_{-0.01}^{+0.01}$ \\
      (CIE comp.) & $Z_{\rm Fe} = Z_{\rm Ni}$ (solar) & ------------ & 1 (fixed) & 1 (fixed) & 1 (fixed) & 1.3 $_{-0.1}^{+0.1}$ \\
      & $Z_{\rm other}$ (solar) & ------------ & 1 (fixed) & 1 (fixed) & 1 (fixed) & 1 (fixed) \\
      & $VEM$ ($10^{57}~\rm cm^{-3}$)$^{\dagger}$ & ------------ & 8.9 $_{-0.6}^{+2.1}$ & 14.2 $_{-1.9}^{+1.6}$ & 3.5 $_{-1.0}^{+1.3}$ & 35.1 $_{-3.9}^{+2.2}$ \\
      VVRNEI2 & $kT_e$ (keV) & 0.30 $_{-0.01}^{+0.01}$ & 0.26 $_{-0.01}^{+0.01}$ & 0.19 $_{-0.01}^{+0.01}$ & 0.22 $_{-0.01}^{+0.06}$ & 0.24 $_{-0.01}^{+0.01}$ \\
      ($\rm RP_{cold}$ comp.)& $kT_{\rm init}$ (keV) & 5 (fixed) & 5 (fixed) & 5 (fixed) & 5 (fixed) & 5 (fixed) \\
      & $Z_{\rm O}$ (solar) & 1 (fixed) & 1 (fixed) & 1 (fixed) & 1 (fixed) & 2.4 $_{-0.3}^{+0.6}$ \\
      & $Z_{\rm Ne}$ (solar) & 1 (fixed) & 2.4 $_{-0.1}^{+0.2}$ & 3.2 $_{-0.2}^{+0.2}$ & 1.5 $_{-0.2}^{+0.2}$ & 3.8 $_{-0.3}^{+0.4}$ \\
      & $Z_{\rm Mg}$ (solar) & 1 (fixed) & 1.5 $_{-0.1}^{+0.1}$ & 1.6 $_{-0.1}^{+0.1}$ & 1.1 $_{-0.1}^{+0.1}$ & 2.2 $_{-0.1}^{+0.3}$ \\
      & $Z_{\rm Si}$ (solar) & 1.1 $_{-0.1}^{+0.1}$ & 1.5 $_{-0.1}^{+0.1}$ & 1.8 $_{-0.1}^{+0.1}$ & 0.8 $_{-0.1}^{+0.1}$ & 3.9 $_{-0.2}^{+0.4}$ \\
      & $Z_{\rm S}$ (solar) & 0.6 $_{-0.1}^{+0.1}$ & 0.9 $_{-0.1}^{+0.1}$ & 1.2 $_{-0.1}^{+0.1}$ & 0.5 $_{-0.1}^{+0.1}$ & 2.6 $_{-0.1}^{+0.4}$ \\
      & $Z_{\rm Ar}$ (solar) & 0.6 $_{-0.1}^{+0.1}$ & 0.8 $_{-0.1}^{+0.1}$ & 1.2 $_{-0.1}^{+0.1}$ & $= Z_{\rm S}$ & 2.4 $_{-0.2}^{+0.2}$ \\
      & $Z_{\rm Ca}$ (solar) & 0.6 $_{-0.1}^{+0.1}$ & 0.8 $_{-0.1}^{+0.2}$ & 1.2 $_{-0.2}^{+0.2}$ & $= Z_{\rm S}$ & 2.2 $_{-0.2}^{+0.3}$ \\
      & $Z_{\rm Fe} = Z_{\rm Ni}$ (solar) & 0.2 $_{-0.1}^{+0.1}$ & 0.1 $_{-0.1}^{+0.1}$ & 0.5 $_{-0.1}^{+0.1}$ & 0.2 $_{-0.1}^{+0.1}$ & 0.8 $_{-0.1}^{+0.1}$ \\
      & $n_{e}t$ (10$^{11}$ s~cm$^{-3}$) & 4.6 $_{-0.1}^{+0.1}$ & 4.1 $_{-0.1}^{+0.2}$ & 4.2 $_{-0.1}^{+0.2}$ & 9.7 $_{-1.1}^{+0.9}$ & 5.4 $_{-0.1}^{+0.1}$ \\
      & $VEM$ ($10^{57}~\rm cm^{-3}$)$^{\dagger}$ & 30.5 $_{-1.4}^{+1.4}$ & 21.7 $_{-0.9}^{+0.5}$ & 12.6 $_{-1.0}^{+0.7}$ & 2.4 $_{-1.4}^{+0.8}$ & 7.0 $_{-0.5}^{+0.4}$ \\
      VVRNEI3 & $kT_e$ (keV) & ------------ & ------------ & 0.54 $_{-0.01}^{+0.03}$ & 0.63 $_{-0.01}^{+0.02}$ & 0.61 $_{-0.01}^{+0.01}$ \\
      ($\rm RP_{hot}$ comp.) & $kT_{\rm init}$ (keV) & ------------ & ------------ & 5 (fixed) & 5 (fixed)  & 5 (fixed) \\
      & $Z_{\rm all}$ (solar) & ------------ & ------------ & = $\rm RP_{cold}$ & = $\rm RP_{cold}$ & = $\rm RP_{cold}$ \\
      & $n_{e}t$ (10$^{11}$ s~cm$^{-3}$) & ------------ & ------------ & = $\rm RP_{cold}$ & = $\rm RP_{cold}$ & = $\rm RP_{cold}$ \\
      & $VEM$ ($10^{57}~\rm cm^{-3}$)$^{\dagger}$ & ------------ & ------------ & 4.1 $_{-0.4}^{+0.2}$ & 1.8 $_{-0.2}^{+0.1}$ & 6.8 $_{-0.9}^{+0.4}$ \\
      Gaussian & Centroid (keV) & ------------ & 1.23 (fixed) & 1.23 (fixed) & 1.23 (fixed) & 1.23 (fixed) \\
      & Normalization$^{\ddagger}$ & ------------ & 132 $_{-15}^{+15}$ & 56 $_{-9}^{+8}$ & 49 $_{-6}^{+16}$ & 142 $_{-19}^{+18}$ \\
      Power law & Photon index & 1.89 (fixed) & 1.89 (fixed) & 1.89 (fixed) & ------------ & ------------ \\
      & Normalization$^{\ast}$ & 151 $_{-23}^{+23}$ & 171 $_{-20}^{+22}$ & 130 $_{-14}^{+12}$ & ------------ & ------------  \\
      \hline
      & $\chi^{2}_{\nu}$ ($\nu$) & 1.55 (594) & 1.71 (1126) & 1.47 (1124) & 1.23 (838) & 1.67 (1123) \\
\enddata
\tablecomments{
$^{\dagger}$Volume emission measure at the distance of 1.5 kpc: $VEM = \int n_e n_{\rm H} dV$ , where $n_e$, $n_{\rm H}$, and $V$ are the electron and hydrogen densities, and the emitting volume, respectively.\\
$^{\ddagger}$The unit is photons s$^{-1}$ cm$^{-2}$ sr$^{-1}$.
$^{\ast}$The unit is photons s$^{-1}$ cm$^{-2}$ keV$^{-1}$ sr$^{-1}$ at 1~keV.\\
}
\end{deluxetable*}

\section{Discussion} \label{sec:discuss}
\subsection{Absorption by the Molecular Cloud}
Figure~\ref{fig:fitting_result1}a shows the spatial distribution of the X-ray absorption column densities (hereafter $N_{\rm H, X}$).
Since our analysis covers almost the whole remnant, we can perform a systematic comparison between $N_{\rm H, X}$ and the ambient gas distribution.
Our spectral fits gave high $N_{\rm H, X}$ in regions 1, 2, 5, 6, 13, 14, 16, 17 and 18, which coincide with the locations of the molecular cloud traced by the $^{12}$CO($J =$~1--0) line \citep[e.g.,][]{Xu2011,Yoshiike2017}. 
Our result indicates that the cloud is located in front of the remnant. 
If we subtract the lowest $N_{\rm H, X}$ (region 19) from the highest one (region 14), we obtain $(3.1 \pm 0.6) \times 10^{21}~\rm cm^{-2}$, 
which can be attributed to the absorption by the molecular cloud.

\cite{Yoshiike2017} observed the $\rm ^{12}CO$($J =$~1--0) line with the NANTEN2 telescope, and concluded that the molecular cloud is located in front of IC~443 based on line velocity measurements. 
We note that \cite{Troja2006} also reached the same conclusion based on a comparison between {\it XMM-Newton} data and CO data. 
\cite{Yoshiike2017} estimated the column density of the molecular cloud to be $N_{\rm H, CO} = (6.1 \pm 0.8) \times 10^{21}~\rm cm^{-2}$. 
Although the estimate by \cite{Yoshiike2017} is somewhat larger than ours, the two measurements  can be regarded to be consistent with each other, considering the fact that the estimate by \cite{Yoshiike2017} is obtained for the CO emission peak location and that our estimate is an average of a much larger region.

\subsection{Origin of the CIE plasma}
\begin{figure*}
\begin{center}
 \includegraphics[width=15cm]{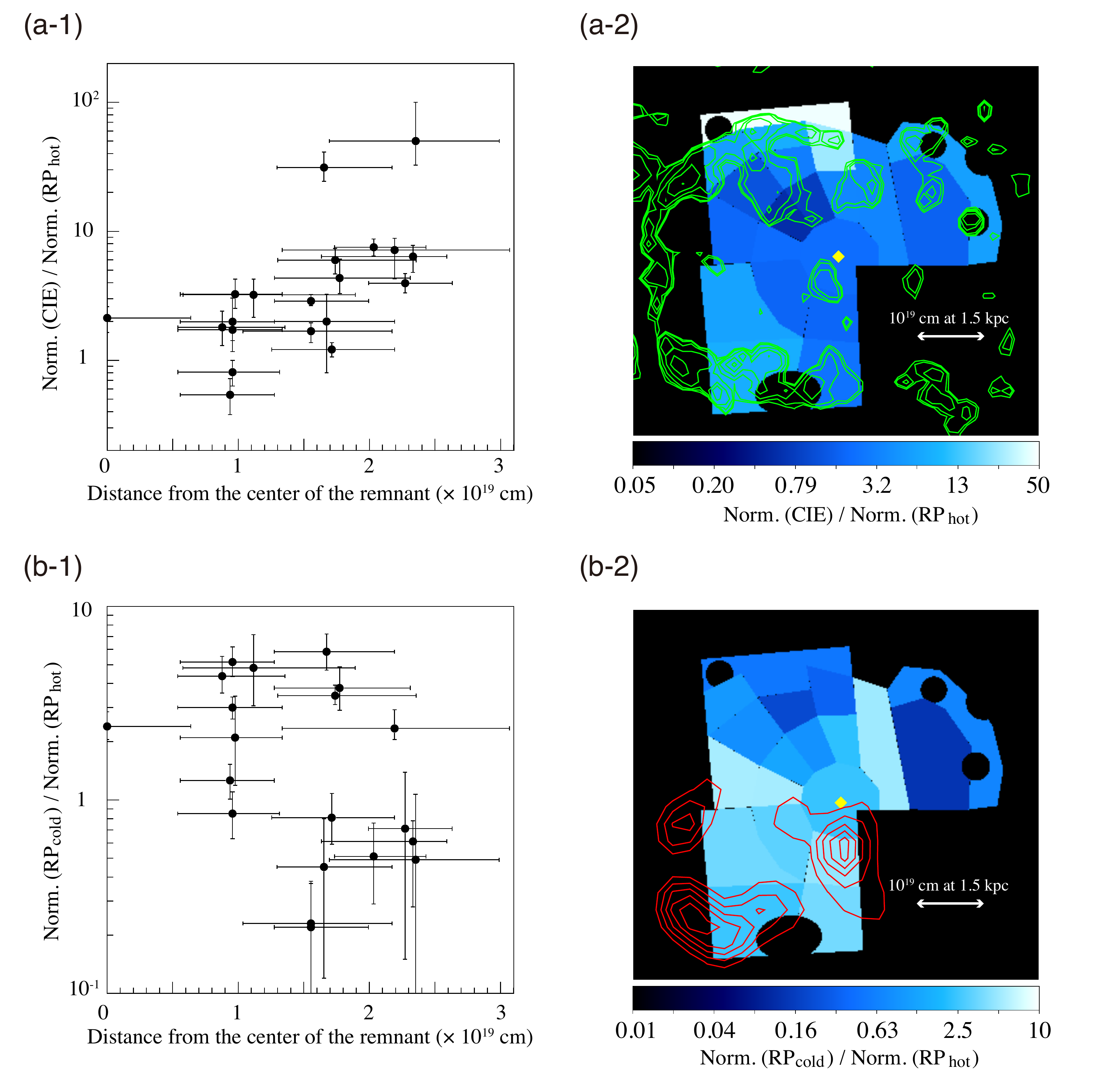} 
\end{center}
\caption{(a-1) Radial profile and (a-2) map of the CIE to $\rm RP_{hot}$ normalization ratio with the 1.4~GHz radio emissions (green contours).
(b-1) Radial profile and (b-2) map of the $\rm RP_{cold}$ to $\rm RP_{hot}$ normalization ratio with 
a map of the $^{12}$CO($J = 2-1$) to $^{12}$CO($J = 1-0$) intensity ratio \citep[red contours;][]{Yoshiike2017}.
The yellow diamonds denote the center of the remnant (R.A.,~Dec.) $=$ ($6^{\rm{h}} 16^{\rm{m}} 40\fs0,~22\degr 38\arcmin 00\farcs$0).
The length of the white arrows correspond to $10^{19}~\rm cm$ at the distance of 1.5~kpc.
}
\label{fig:normratio}
\end{figure*}

In the spectral analysis, we successfully reproduced the spectra from all the regions with the model consisting of 
the CIE, $\rm RP_{cold}$ and $\rm RP_{hot}$ components.
The abundances of the CIE component are consistent with solar, suggesting that the component is of shocked ISM origin. 
The spatial distribution of the component also supports this picture. 
Figure~\ref{fig:normratio}a-1 shows the normalization ratios of the CIE component to the $\rm RP_{hot}$ component plotted as a function of the angular distance from the center of the remnant which we define as (R.A.,~Dec.) $=$ ($6^{\rm{h}} 16^{\rm{m}} 40\fs0,~22\degr 38\arcmin 00\farcs$0). 
The map of the normalization ratio is shown in Figure~\ref{fig:normratio}a-2. 
The normalization ratios tend to increase toward the outer regions, suggesting that the CIE plasma is concentrated in the rim as is generally expected for a shocked ISM emission.

\subsection{Origin of the recombining plasmas}
In section~\ref{sec:alaysis_snr}, we found that the emission is well reproduced by a model which includes two RP components with different $kT_e$.
As shown in Figure~\ref{fig:kTe}, the electron temperatures of both components decrease toward the southeast.
The southeastern region coincides with the locations where the molecular cloud is known to be interacting with the remnant based on the $^{12}$CO($J =$~2--1) to $^{12}$CO($J =$~1--0) intensity ratio \citep[e.g.,][]{Xu2011,Yoshiike2017}.
The normalization ratio of the two RP components points to a similar tendency. 
Figures~\ref{fig:normratio}b-1 and \ref{fig:normratio}b-2 show a radial profile and a map of the $\rm RP_{cold}$ to $\rm RP_{hot}$ 
normalization ratio, respectively. 
As opposed to Figure~\ref{fig:normratio}a-1, no clear correlation is seen in the radial profile. 
From the map, it is evident that the ratio is larger in the southeastern region, where $kT_e$ is also lower than the other regions (see Figure~\ref{fig:kTe}). 
These results on $kT_e$ and the normalization ratio together indicate that the SNR plasma is cooler in the region where the shock 
is in contact with the molecular cloud. 

One of the plausible mechanisms to explain the cooler plasma in the southeastern region is cooling of the X-ray emitting plasma by the molecular cloud via thermal conduction. 
If the cooling proceeds faster than the recombination timescale of the plasma, the recombining plasma can also be explained by the thermal conduction. 
In this scenario, continuous cooling until the present day may have resulted in the smallest $n_e t$ in the southeast (Figure~\ref{fig:fitting_result1}b). 
The thermal conduction scenario is proposed for the formation process of an RP in G166.0+4.3 by \cite{Matsumura2017}, who 
found $kT_e$ of a high-density region of the SNR is significantly lower 
than that of a low-density region. 
Their observational result is in fact similar to our findings on IC~443 reported in this paper. 
On the contrary, as discussed by \cite{Lopez2013} for W49B, higher $kT_e$ would be expected in a region with a high ambient gas density if we 
consider rarefaction as the formation process of RPs. 

It is interesting to point out that the plasma in the northwestern region is close to CIE with $n_et \sim 10^{12}~\rm s~cm^{-3}$.  
A possible explanation is that the cooling of the plasma in this region is less efficient since the region is far away from the molecular cloud, 
which would be the major cooling source for IC~443. 
Lower ambient gas density, which is suggested by the large shell radius in the west \citep{Lee2008}, would also make the cooling less 
efficient.
~~~\\

\section{Conclusions} \label{sec:conclusion}
We have performed spatially resolved spectroscopy of X-ray emission from IC~443 with {\it Suzaku}. 
Spectra extracted from each region of the remnant are all fitted well with a model consisting of CIE and two RP components. 
The X-ray absorption column densities are high in regions where the $\rm ^{12}CO$($J =$~1--0) emission is bright, indicating that the molecular cloud is located in front of the remnant. 
The metal abundances of the CIE component as well as its spatial distribution suggest that it is emitted by the shocked ISM. 
The electron temperature of one of the RP component ranges from 0.16~keV to 0.28~keV whereas that of the other RP component is in the range of 0.48--0.67~keV.
The electron temperatures of both RP components decrease toward the southeast, where the remnant is interacting with the molecular cloud.
Also, the normalization ratio of the lower-$kT_e$ RP to higher-$kT_e$ RP components increases toward the southeast. 
The two findings may be a result of cooling of the X-ray emitting plasma by the interacting molecular cloud via thermal conduction. 
The thermal conduction can explain the RP if the cooling proceeds fast enough compared to the recombination timescale of the plasma. 

\acknowledgments

We thank Prof. Katsuji Koyama and Dr. Hiroya Yamaguchi for helpful advice.
We are grateful to Dr. Satoshi Yoshiike for providing us with the NANTEN2 data.
We deeply appreciate all the {\it Suzaku} team members. 
This work is partially supported by JSPS/MEXT Scientific Research Grant Numbers JP15J01842 (H.M.), JP25109004 (T.T. and T.G.T.), JP26800102 (H.U.), JP15H02090 (T.G.T.), and JP26610047 (T.G.T.).



\appendix

\section{Fitting results of the spectra of the regions A and B} \label{apx:point}
\begin{figure*}[h]
\begin{center}
 \includegraphics[width=12cm]{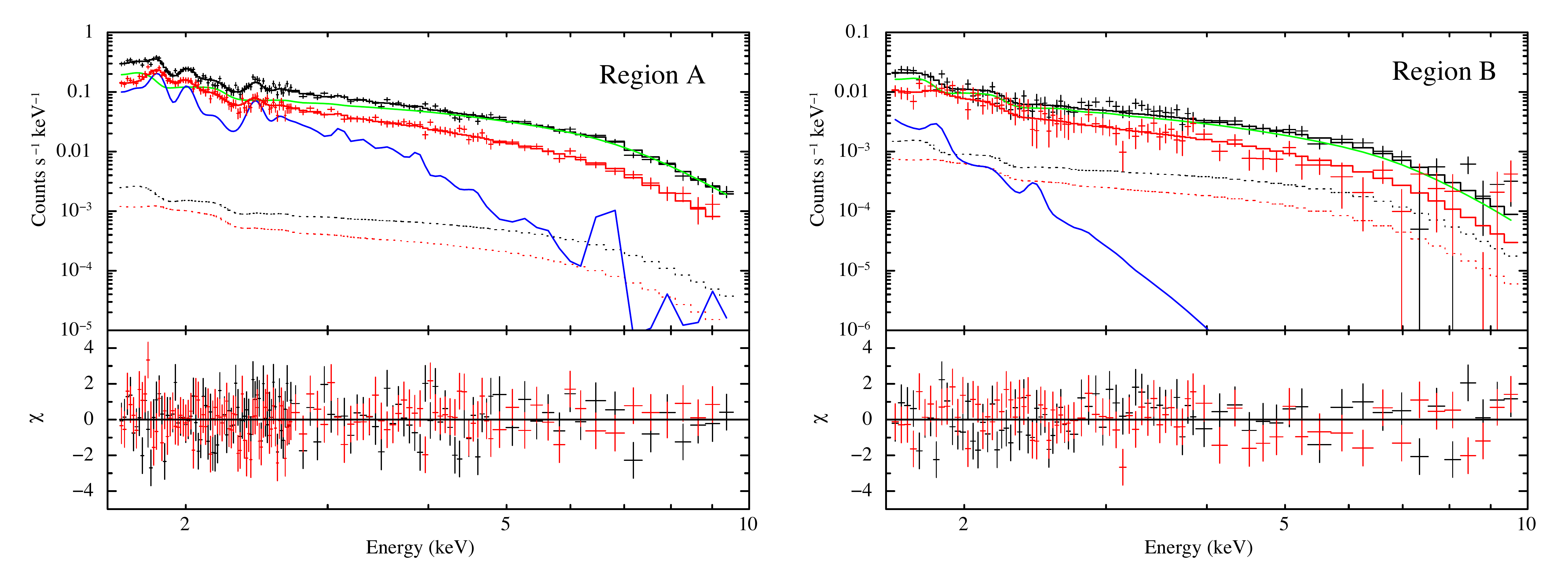} 
\end{center}
\caption{
Same as Figure~\ref{fig:SE_spe} but for regions A and B.
}
\label{fig:pointing_sources_spe}
\end{figure*}

Figure~\ref{fig:pointing_sources_spe} shows spectra of the PWN 1SAX J0617.1+2221 in region~A and the point source U061530.75+224910.6 in region~B in the 1.6-10.0~keV band after NXB subtraction.
We fitted the spectra with a model consisting of a power law and an RP component for the sources and the SNR plasma, respectively.
The X-ray absorption was fixed at $7.0 \times 10^{21}~\rm cm^{-2}$ as determined by \cite{Kawasaki2002}.
We fixed $kT_{\rm init}$ at 5~keV, whereas $kT_e$ and $n_et$ were allowed to vary.
The abundances of Si and S were allowed to vary, whereas those of Ar and Ca were linked to S.
The abundances of the other elements were fixed to solar.
The spectra can be reproduced well by the model (Table~\ref{tab:pointing_model}).
The photon indices of PWN 1SAX J0617.1+2221 and U061530.75+224910.6 are $1.89_{-0.07}^{+0.08}$ and $2.22_{-0.16}^{+0.15}$, respectively, which we used in the SNR emission analysis (see subsection~\ref{sec:alaysis_snr}).
The photon indices and the fluxes of the power-law components are consistent with the results by \cite{Bocchino2003}, who analyzed {\it XMM-Newton} data. 

\begin{deluxetable}{cccc}[h]
\tablecaption{Best-fit model parameters of the spectra in regions A and B.
\label{tab:pointing_model}}
\tablehead{
\colhead{} & 
\colhead{} & 
\colhead{1SAX J0617.1+2221} & 
\colhead{U061530.75+224910.6} \\
\colhead{Model function} & 
\colhead{Parameter} &
\colhead{region~A} &
\colhead{region~B}
} 
\startdata
      TBabs & $N_{\rm H}$ (10$^{22}$ cm$^{-2}$) & 0.7 (fixed) & 0.7 (fixed) \\
      Power law & Photon index & 1.89 $_{-0.07}^{+0.08}$ & 2.22 $_{-0.16}^{+0.15}$ \\
      & Normalization$^{\ast}$ & 801 $_{-94}^{+114}$ & 254 $_{-50}^{+30}$ \\
      VVRNEI & $kT_e$ (keV) & 0.34 $_{-0.04}^{+0.05}$ & 0.31 $_{-0.20}^{+0.49}$ \\
      (RP comp.) & $kT_{init}$ (keV) & 5 (fixed) & 5 (fixed) \\
      & $Z_{\rm Si}$ (solar) & 2.4 $_{-0.5}^{+0.7}$ & 1 (fixed) \\
      & $Z_{\rm S}$ = $Z_{\rm Ar}$ = $Z_{\rm Ca}$ (solar) &  1.3 $_{-0.3}^{+0.4}$ & 1 (fixed) \\
      & $n_{e}t$ (10$^{11}$ s~cm$^{-3}$) & 4.2 $_{-0.4}^{+0.6}$ & $>100$ \\
      & $VEM$ ($10^{57}~\rm cm^{-3}$)$^{\dagger}$ & 1.2 $_{-0.4}^{+0.2}$ & 1.0 $_{-1.0}^{+5.7}$ \\
      \hline
      & $\chi^{2}_{\nu}$ ($\nu$) & 1.16 (241) & 1.05 (149) \\
\enddata
\tablecomments{
$^{\ast}$The unit is photons s$^{-1}$cm$^{-2}$keV$^{-1}$sr$^{-1}$ @ 1~keV.\\
$^{\dagger}$Volume emission measure at the distance of 1.5 kpc: $VEM = \int n_e n_{\rm H} dV$ , where $n_e$, $n_{\rm H}$, and $V$ are the electron and hydrogen densities, and the emitting volume, respectively.
}
\end{deluxetable}

\section{Fitting results of the spectra from regions 1--20} \label{apx:1-20}

\begin{figure*}[h]
\begin{center}
 \includegraphics[width=16cm]{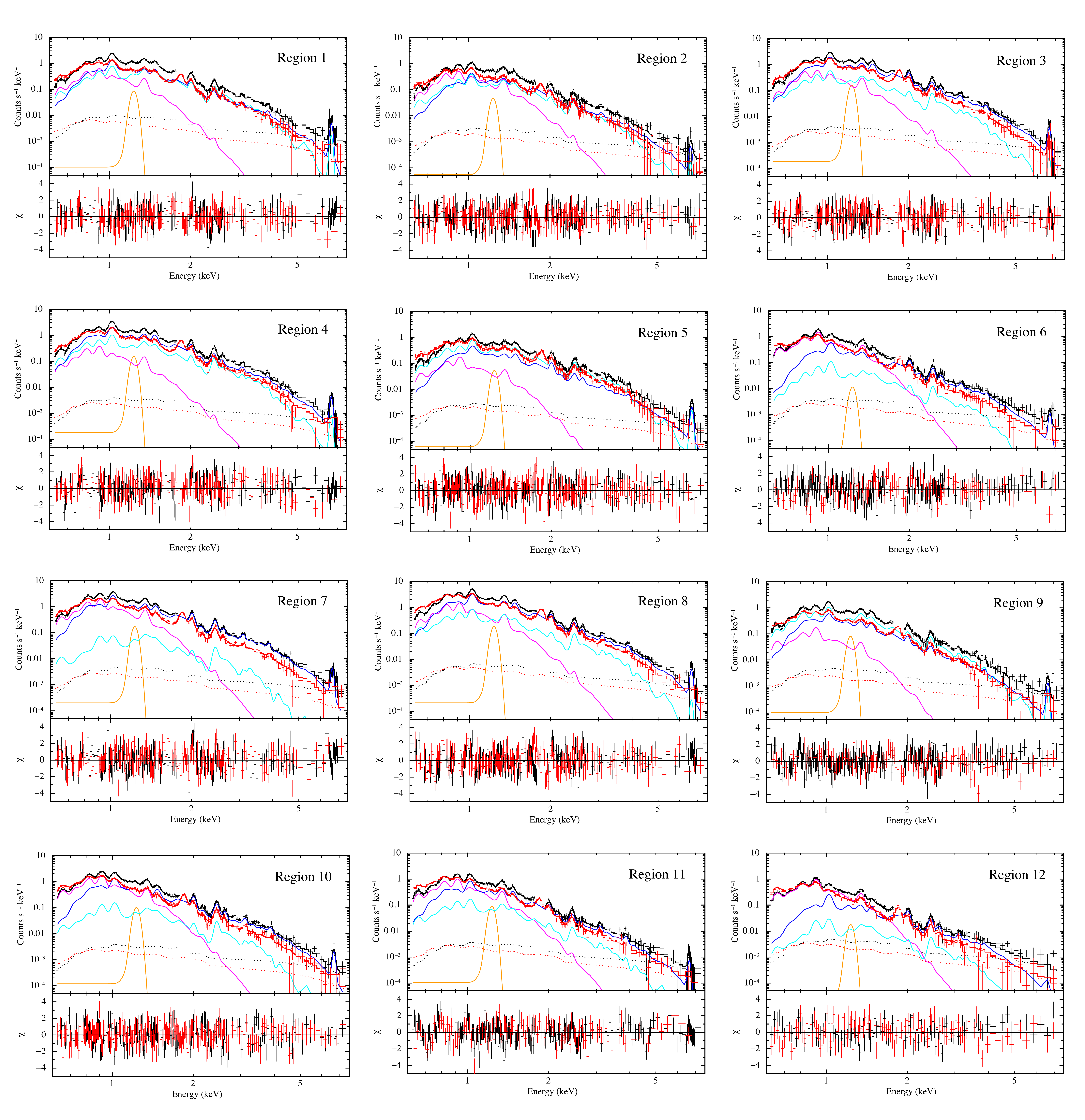} 
\end{center}
\caption{
Same as Figure~\ref{fig:SE_spe} but for regions 1--12.
}
\label{fig:spe1to12}
\end{figure*}

\begin{figure*}[t]
\begin{center}
 \includegraphics[width=16cm]{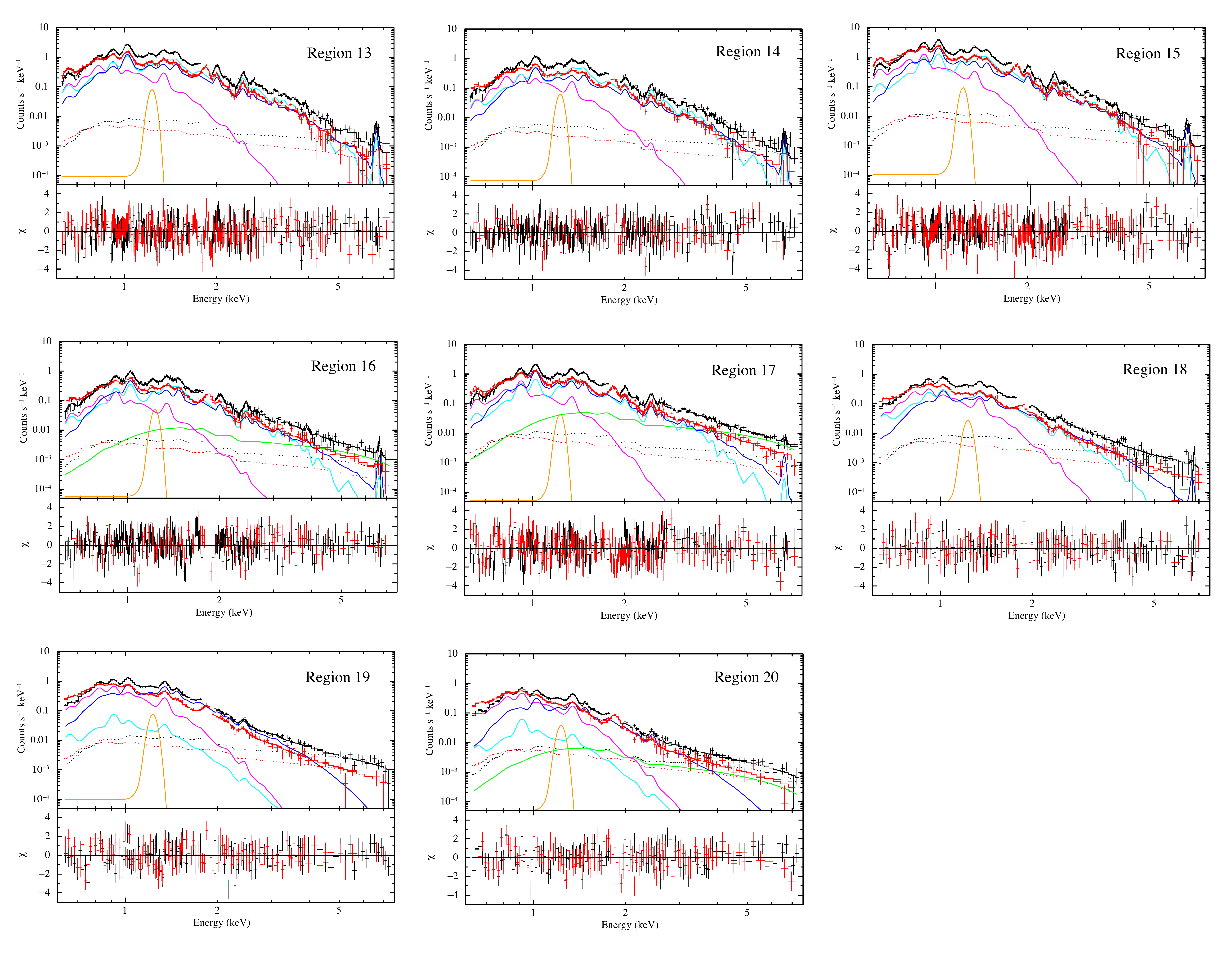} 
\end{center}
\caption{Same as Figure~\ref{fig:SE_spe} but for regions 13--20.}
\label{fig:spe13to20}
\end{figure*}

\begin{longrotatetable}
\begin{deluxetable*}{cccccccccccc}
\tablecaption{Best-fit model parameters of the spectra from regions 1--10.
\label{tab:reg1to10_model}}
\tablewidth{700pt}
\tabletypesize{\scriptsize}
\tablehead{
\colhead{Model function} & 
\colhead{Parameter} &
\colhead{reg. 1} &
\colhead{reg. 2} &
\colhead{reg. 3} &
\colhead{reg. 4} &
\colhead{reg. 5} &
\colhead{reg. 6} &
\colhead{reg. 7} &
\colhead{reg. 8} &
\colhead{reg. 9} &
\colhead{reg. 10}
} 
\startdata
      TBabs & $N_{\rm H}$ (10$^{22}$ cm$^{-2}$) & 0.94 $_{-0.04}^{+0.04}$ & 0.95 $_{-0.04}^{+0.04}$ & 0.82 $_{-0.04}^{+0.03}$ & 0.79 $_{-0.03}^{+0.03}$ & 0.88 $_{-0.06}^{+0.05}$ & 0.89 $_{-0.02}^{+0.02}$ & 0.79 $_{-0.02}^{+0.02}$ & 0.70 $_{-0.02}^{+0.03}$ & 0.84 $_{-0.05}^{+0.04}$ & 0.74 $_{-0.03}^{+0.02}$ \\
      VVRNEI1 & $kT_e$ (keV) & 0.24 $_{-0.01}^{+0.02}$ & 0.26 $_{-0.02}^{+0.01}$ & 0.28 $_{-0.01}^{+0.05}$ & 0.25 $_{-0.01}^{+0.04}$& 0.22 $_{-0.03}^{+0.05}$ & 0.21 $_{-0.01}^{+0.01}$ & 0.24 $_{-0.01}^{+0.01}$ & 0.26 $_{-0.01}^{+0.01}$ & 0.20 $_{-0.06}^{+0.06}$ & 0.23 $_{-0.01}^{+0.01}$ \\
       (CIE comp.) & $Z_{\rm all}$ (solar) & 1 (fixed) & 1 (fixed) & 1 (fixed) & 1 (fixed) & 1 (fixed) & 1 (fixed) & 1 (fixed) & 1 (fixed) & 1 (fixed) & 1 (fixed) \\
       & $VEM$ ($10^{57}~\rm cm^{-3}$)$^{\dagger}$ & 2.9 $_{-0.9}^{+1.3}$ & 1.7 $_{-0.5}^{+0.6}$ & 1.1 $_{-0.4}^{+0.5}$ & 0.8 $_{-0.4}^{+0.4}$ & 0.8 $_{-0.3}^{+0.7}$ & 7.5 $_{-0.9}^{+1.1}$ & 4.3 $_{-0.4}^{+1.7}$ & 2.7 $_{-0.5}^{+0.6}$ & 1.1 $_{-1.0}^{+1.1}$ & 4.5 $_{-0.7}^{+1.0}$ \\
      VVRNEI2 & $kT_e$ (keV) & 0.22 $_{-0.01}^{+0.01}$ & 0.28 $_{-0.07}^{+0.05}$ & 0.25 $_{-0.05}^{+0.05}$ & 0.27 $_{-0.02}^{+0.02}$ & 0.25 $_{-0.01}^{+0.02}$ & 0.27 $_{-0.09}^{+0.07}$ & 0.16 $_{-0.07}^{+0.08}$ & 0.26 $_{-0.01}^{+0.01}$ & 0.25 $_{-0.05}^{+0.04}$ & 0.22 $_{-0.03}^{+0.07}$ \\
       ($\rm RP_{cold}$ comp.) & $kT_{\rm init}$ (keV) & 5 (fixed) & 5 (fixed) & 5 (fixed) & 5 (fixed) & 5 (fixed) & 5 (fixed) & 5 (fixed) & 5 (fixed) & 5 (fixed) & 5 (fixed) \\
       & $Z_{\rm Ne}$ (solar) & 2.5 $_{-0.3}^{+0.4}$ & 1.9 $_{-0.4}^{+0.3}$ & 2.2 $_{-0.3}^{+0.2}$ & 1.9 $_{-0.1}^{+0.2}$ & 2.3 $_{-0.3}^{+0.3}$ & 5.1 $_{-1.5}^{+1.7}$ & 2.2 $_{-0.2}^{+0.3}$ & 1.8 $_{-0.1}^{+0.1}$ & 2.0 $_{-0.3}^{+0.2}$ & 3.4 $_{-0.5}^{+0.5}$ \\
       & $Z_{\rm Mg}$ (solar) & 1.5 $_{-0.2}^{+0.1}$ & 1.2 $_{-0.2}^{+0.2}$ & 1.5 $_{-0.2}^{+0.1}$ & 1.4 $_{-0.1}^{+0.1}$ & 1.3 $_{-0.1}^{+0.3}$ & 2.2 $_{-0.6}^{+0.8}$ & 1.3 $_{-0.1}^{+0.2}$ & 1.4 $_{-0.1}^{+0.1}$ & 1.4 $_{-0.1}^{+0.2}$ & 1.7 $_{-0.3}^{+0.3}$ \\
       & $Z_{\rm Si}$ (solar) & 2.0 $_{-0.2}^{+0.2}$ & 1.7 $_{-0.3}^{+0.2}$ & 2.7 $_{-0.2}^{+0.2}$ & 2.5 $_{-0.1}^{+0.2}$ & 2.4 $_{-0.2}^{+0.2}$ & 4.9 $_{-1.1}^{+1.6}$ & 3.4 $_{-0.3}^{+0.1}$ & 2.2 $_{-0.2}^{+0.1}$ & 1.5 $_{-0.1}^{+0.2}$ & 4.2 $_{-0.5}^{+0.6}$ \\
       & $Z_{\rm S}$ (solar) & 1.2 $_{-0.2}^{+0.1}$ & 1.0 $_{-0.2}^{+0.2}$ & 1.7 $_{-0.1}^{+0.2}$ & 1.6 $_{-0.2}^{+0.1}$ & 1.4 $_{-0.1}^{+0.2}$ & 3.6 $_{-0.9}^{+0.9}$ & 2.4 $_{-0.2}^{+0.3}$ & 1.6 $_{-0.2}^{+0.1}$ & 0.9 $_{-0.1}^{+0.1}$ & 3.3 $_{-0.4}^{+0.4}$ \\
       & $Z_{\rm Ar} = Z_{\rm Ca}$ (solar) & 1.3 $_{-0.2}^{+0.3}$ & 0.9 $_{-0.4}^{+0.5}$ & 1.6 $_{-0.2}^{+0.3}$ & 1.4 $_{-0.3}^{+0.2}$ & 1.3 $_{-0.3}^{+0.3}$ & 2.4 $_{-1.2}^{+1.2}$ & 2.5 $_{-0.4}^{+0.4}$ & 1.9 $_{-0.2}^{+0.3}$ & 1.0 $_{-0.2}^{+0.3}$ & 3.1 $_{-0.6}^{+0.8}$ \\
       & $Z_{\rm Fe} = Z_{\rm Ni}$ (solar) & 0.7 $_{-0.1}^{+0.1}$ & 0.4 $_{-0.1}^{+0.1}$ & 0.5 $_{-0.1}^{+0.1}$ & 0.5 $_{-0.1}^{+0.1}$ & 0.8 $_{-0.2}^{+0.1}$ & 1.1 $_{-0.3}^{+0.5}$ & 0.6 $_{-0.1}^{+0.1}$ & 0.5 $_{-0.1}^{+0.1}$ & 0.8 $_{-0.1}^{+0.2}$ & 0.8 $_{-0.1}^{+0.2}$ \\
       & $n_{e}t$ (10$^{11}$ s~cm$^{-3}$) & 4.1 $_{-0.2}^{+0.3}$ & 5.6 $_{-0.6}^{+0.6}$ & 5.4 $_{-0.2}^{+0.4}$ & 5.1 $_{-0.2}^{+0.2}$ & 4.4 $_{-0.3}^{+0.2}$ & 5.6 $_{-0.4}^{+0.5}$ & 5.4 $_{-0.3}^{+0.1}$ & 5.7 $_{-0.2}^{+0.2}$ & 5.4 $_{-0.3}^{+0.4}$ & 5.6 $_{-0.3}^{+0.2}$ \\
       & $VEM$ ($10^{57}~\rm cm^{-3}$)$^{\dagger}$ & 3.3 $_{-0.5}^{+0.5}$ & 1.1 $_{-0.5}^{+0.7}$ & 1.2 $_{-0.3}^{+0.4}$ & 1.8 $_{-0.4}^{+0.3}$ & 2.1 $_{-0.3}^{+0.3}$ & 0.1 $_{-0.1}^{+0.1}$ & 0.3 $_{-0.2}^{+0.2}$ & 1.8 $_{-0.5}^{+0.6}$ & 3.2 $_{-0.5}^{+0.5}$ & 0.3 $_{-0.1}^{+0.1}$ \\
      VVRNEI3 & $kT_e$ (keV) & 0.56 $_{-0.05}^{+0.04}$ & 0.62 $_{-0.07}^{+0.07}$ & 0.62 $_{-0.03}^{+0.02}$ & 0.59 $_{-0.02}^{+0.03}$ & 0.67 $_{-0.05}^{+0.06}$ & 0.64 $_{-0.04}^{+0.06}$ & 0.55 $_{-0.02}^{+0.03}$ & 0.57 $_{-0.01}^{+0.02}$ & 0.61 $_{-0.06}^{+0.04}$ & 0.61 $_{-0.03}^{+0.03}$ \\
       ($\rm RP_{hot}$ comp.) & $kT_{\rm init}$ (keV) & 5 (fixed) & 5 (fixed) & 5 (fixed) & 5 (fixed) & 5 (fixed) & 5 (fixed) & 5 (fixed) & 5 (fixed) & 5 (fixed) & 5 (fixed) \\
       & $Z_{\rm all}$ (solar) & = $\rm RP_{cold}$ & = $\rm RP_{cold}$ & = $\rm RP_{cold}$ & = $\rm RP_{cold}$ & = $\rm RP_{cold}$ & = $\rm RP_{cold}$ & = $\rm RP_{cold}$ & = $\rm RP_{cold}$ & = $\rm RP_{cold}$ & = $\rm RP_{cold}$ \\
       & $n_{e}t$ (10$^{11}$ s~cm$^{-3}$) & = $\rm RP_{cold}$ & = $\rm RP_{cold}$ & = $\rm RP_{cold}$ & = $\rm RP_{cold}$ & = $\rm RP_{cold}$ & = $\rm RP_{cold}$ & = $\rm RP_{cold}$ & = $\rm RP_{cold}$ & = $\rm RP_{cold}$ & = $\rm RP_{cold}$ \\
       & $VEM$ ($10^{57}~\rm cm^{-3}$)$^{\dagger}$ & 1.4 $_{-0.3}^{+0.2}$ & 0.5 $_{-0.2}^{+0.3}$ & 1.4 $_{-0.2}^{+0.2}$ & 1.4 $_{-0.2}^{+0.2}$ & 0.4 $_{-0.1}^{+0.1}$ & 0.3 $_{-0.1}^{+0.1}$ & 1.5 $_{-0.1}^{+0.2}$ & 2.2 $_{-0.2}^{+0.2}$ & 0.6 $_{-0.1}^{+0.1}$ & 0.6 $_{-0.1}^{+0.1}$ \\
      Gaussian & Centroid (keV) & 1.23 (fixed) & 1.23 (fixed) & 1.23 (fixed) & 1.23 (fixed) & 1.23 (fixed) & 1.23 (fixed) & 1.23 (fixed) & 1.23 (fixed) & 1.23 (fixed) & 1.23 (fixed) \\
       & Normalization$^{\ddagger}$ & 14 $_{-7}^{+7}$ & 19 $_{-11}^{+11}$ & 47 $_{-13}^{+10}$ & 46 $_{-12}^{+13}$ & 24 $_{-6}^{+13}$ & 3 $_{-3}^{+9}$ & 42 $_{-12}^{+11}$ & 30 $_{-8}^{+8}$ & 22 $_{-8}^{+8}$ & 30 $_{-12}^{+10}$ \\
      Power law & Photon index & ------------ & ------------ & ------------ & ------------ & ------------ & ------------ & ------------ & ------------ & ------------ & ------------ \\
       & Normalization$^{\ast}$ & ------------ & ------------ & ------------ & ------------ & ------------ & ------------ & ------------ & ------------ & ------------ & ------------ \\
      \hline
      & $\chi^{2}_{\nu}$ ($\nu$) & 1.13 (477) & 1.15 (477) & 1.12 (485) & 1.32 (477) & 1.17 (477) & 1.24 (365) & 1.39 (477) & 1.30 (477) & 1.04 (365) & 1.16 (477) \\
\enddata
\tablecomments{
$^{\dagger}$Volume emission measure at the distance of 1.5 kpc: $VEM = \int n_e n_{\rm H} dV$ , where $n_e$, $n_{\rm H}$, and $V$ are the electron and hydrogen densities, and the emitting volume, respectively.\\
$^{\ddagger}$The unit is photons s$^{-1}$ cm$^{-2}$ sr$^{-1}$.\\
$^{\ast}$The unit is photons s$^{-1}$ cm$^{-2}$ keV$^{-1}$ sr$^{-1}$ at 1~keV.\\
}
\end{deluxetable*}
\end{longrotatetable}

\begin{longrotatetable}
\begin{deluxetable*}{cccccccccccc}
\tablecaption{Best-fit model parameters of the spectra from regions 11--20.
\label{tab:reg11to20_model}}
\tablewidth{700pt}
\tabletypesize{\scriptsize}
\tablehead{
\colhead{Model function} & 
\colhead{Parameter} &
\colhead{reg. 11} &
\colhead{reg. 12} &
\colhead{reg. 13} &
\colhead{reg. 14} &
\colhead{reg. 15} &
\colhead{reg. 16} &
\colhead{reg. 17} &
\colhead{reg. 18} &
\colhead{reg. 19} &
\colhead{reg. 20}
} 
\startdata
      TBabs & $N_{\rm H}$ (10$^{22}$ cm$^{-2}$) & 0.74 $_{-0.03}^{+0.04}$ & 0.81 $_{-0.03}^{+0.02}$ & 0.88 $_{-0.03}^{+0.03}$ & 0.98 $_{-0.05}^{+0.04}$ & 0.77 $_{-0.03}^{+0.03}$ & 0.91 $_{-0.09}^{+0.14}$ & 0.96 $_{-0.03}^{+0.03}$ & 0.85 $_{-0.03}^{+0.04}$ & 0.67 $_{-0.03}^{+0.02}$ & 0.77 $_{-0.02}^{+0.05}$ \\
      VVRNEI & $kT_e$ (keV) & 0.24 $_{-0.01}^{+0.01}$ & 0.22 $_{-0.01}^{+0.01}$ & 0.24 $_{-0.02}^{+0.02}$ & 0.27 $_{-0.03}^{+0.02}$ & 0.20 $_{-0.01}^{+0.02}$ & 0.24 $_{-0.05}^{+0.05}$ & 0.19 $_{-0.01}^{+0.01}$ & 0.25 $_{-0.06}^{+0.05}$ & 0.27 $_{-0.01}^{+0.01}$ & 0.24 $_{-0.01}^{+0.01}$ \\
       (CIE comp.) & $Z_{\rm all}$ (solar) & 1 (fixed) & 1 (fixed) & 1 (fixed) & 1 (fixed) & 1 (fixed) & 1 (fixed) & 1 (fixed) & 1 (fixed) & 1 (fixed) & 1 (fixed) \\
       & $VEM$ ($10^{57}~\rm cm^{-3}$)$^{\dagger}$ & 2.7 $_{-0.4}^{+0.8}$ & 4.5 $_{-0.7}^{+0.6}$ & 2.1 $_{-0.5}^{+0.6}$ & 0.9 $_{-0.3}^{+0.3}$ & 5.5 $_{-1.5}^{+1.5}$ & 0.7 $_{-0.4}^{+2.0}$ & 5.3 $_{-0.7}^{+1.3}$ & 1.0 $_{-0.4}^{+0.3}$ & 1.3 $_{-0.2}^{+0.2}$ & 1.7 $_{-0.6}^{+0.8}$  \\
      VVRNEI1 & $kT_e$ (keV) & 0.24 $_{-0.07}^{+0.05}$ & 0.26 $_{-0.08}^{+0.09}$ & 0.21 $_{-0.11}^{+0.02}$ & 0.19 $_{-0.01}^{+0.02}$ & 0.18 $_{-0.01}^{+0.01}$ & 0.19 $_{-0.02}^{+0.01}$ & 0.19 $_{-0.03}^{+0.04}$ & 0.22 $_{-0.02}^{+0.02}$ & 0.24 $_{-0.05}^{+0.04}$ & 0.22 $_{-0.08}^{+0.09}$ \\
       ($\rm RP_{cold}$ comp.) & $kT_{\rm init}$ (keV) & 5 (fixed) & 5 (fixed) & 5 (fixed) & 5 (fixed) & 5 (fixed) & 5 (fixed) & 5 (fixed) & 5 (fixed) & 5 (fixed) & 5 (fixed) \\
       & $Z_{\rm Ne}$ (solar) & 2.0 $_{-0.3}^{+0.3}$ & 6.8 $_{-2.9}^{+5.4}$ & 2.3 $_{-0.4}^{+0.3}$ & 2.4 $_{-0.4}^{+0.5}$ & 4.4 $_{-0.6}^{+0.7}$ & 3.9 $_{-0.9}^{+2.0}$ & 5.0 $_{-0.5}^{+0.4}$ & 1.9 $_{-0.3}^{+0.3}$ & 1.5 $_{-0.2}^{+0.1}$ & 2.6 $_{-0.7}^{+1.1}$ \\
       & $Z_{\rm Mg}$ (solar) & 1.5 $_{-0.2}^{+0.3}$ & 2.9 $_{-1.3}^{+4.1}$ & 1.3 $_{-0.3}^{+0.1}$ & 1.4 $_{-0.3}^{+0.2}$ & 1.9 $_{-0.1}^{+0.3}$ & 2.0 $_{-0.5}^{+0.8}$ & 2.4 $_{-0.2}^{+0.2}$ & 1.1 $_{-0.2}^{+0.1}$ & 1.1 $_{-0.1}^{+0.1}$ & 1.5 $_{-0.2}^{+0.4}$ \\
       & $Z_{\rm Si}$ (solar) & 1.7 $_{-0.2}^{+0.4}$ & 9.5 $_{-3.5}^{+5.3}$ & 1.6 $_{-0.2}^{+0.1}$ & 1.2 $_{-0.2}^{+0.2}$ & 2.3 $_{-0.2}^{+0.3}$ & 1.9 $_{-0.3}^{+0.6}$ & 2.5 $_{-0.2}^{+0.2}$ & 0.7 $_{-0.2}^{+0.1}$ & 0.7 $_{-0.1}^{+0.1}$ & 1.4 $_{-0.2}^{+0.4}$ \\
       & $Z_{\rm S}$ (solar) & 1.1 $_{-0.2}^{+0.2}$ & 6.7 $_{-2.5}^{+14.3}$ & 0.9 $_{-0.1}^{+0.1}$ & 0.8 $_{-0.1}^{+0.1}$ & 1.4 $_{-0.2}^{+0.1}$ & 1.1 $_{-0.2}^{+0.4}$ & 1.8 $_{-0.2}^{+0.2}$ & 0.3 $_{-0.1}^{+0.1}$ & 0.4 $_{-0.1}^{+0.1}$ & 0.8 $_{-0.2}^{+0.3}$ \\
       & $Z_{\rm Ar} = Z_{\rm Ca}$ (solar) & 1.8 $_{-0.5}^{+0.6}$ & 5.2 $_{-3.1}^{+5.4}$ & 0.8 $_{-0.2}^{+0.2}$ & 0.8 $_{-0.2}^{+0.2}$ & 1.4 $_{-0.2}^{+0.3}$ & 1.2 $_{-0.5}^{+0.6}$ & 2.0 $_{-0.5}^{+0.5}$ & = $Z_{\rm S}$ & = $Z_{\rm S}$ & = $Z_{\rm S}$ \\
       & $Z_{\rm Fe} = Z_{\rm Ni}$ (solar) & 0.5 $_{-0.1}^{+0.2}$ & 0.8 $_{-0.5}^{+1.2}$ & 0.3 $_{-0.1}^{+0.1}$ & 0.4 $_{-0.2}^{+0.2}$ & 0.8 $_{-0.2}^{+0.2}$ & 0.7 $_{-0.2}^{+0.5}$ & 0.6 $_{-0.2}^{+0.1}$ & 0.3 $_{-0.1}^{+0.2}$ & 0.2 $_{-0.1}^{+0.1}$ & 0.4 $_{-0.1}^{+0.1}$ \\
       & $n_{e}t$ (10$^{11}$ s~cm$^{-3}$) & 5.8 $_{-0.4}^{+0.5}$ & 5.7 $_{-0.5}^{+0.7}$ & 3.8 $_{-0.3}^{+0.3}$ & 3.8 $_{-0.4}^{+0.4}$ & 4.3 $_{-0.1}^{+0.2}$ & 4.7 $_{-0.5}^{+0.4}$ & 4.8 $_{-0.2}^{+0.2}$ & 5.8 $_{-0.7}^{+0.7}$ & 12.3 $_{-1.7}^{+2.3}$ & 15.1 $_{-4.1}^{+8.3}$ \\
       & $VEM$ ($10^{57}~\rm cm^{-3}$)$^{\dagger}$ & 0.5 $_{-0.4}^{+0.4}$ & $\leq$ 0.1 & 3.7 $_{-0.4}^{+0.5}$ & 2.3 $_{-0.5}^{+0.3}$ & 3.2 $_{-0.5}^{+0.6}$ & 1.1 $_{-0.4}^{+0.3}$ & 1.7 $_{-0.8}^{+0.6}$ & 1.5 $_{-0.4}^{+0.6}$ & $\leq$ 0.2 & 0.2 $_{-0.2}^{+0.4}$ \\
      VVRNEI3 & $kT_e$ (keV) & 0.54 $_{-0.08}^{+0.03}$ & 0.63 $_{-0.08}^{+0.10}$ & 0.48 $_{-0.19}^{+0.07}$ & 0.59 $_{-0.14}^{+0.07}$ & 0.54 $_{-0.03}^{+0.04}$ & 0.64 $_{-0.11}^{+0.06}$ & 0.55 $_{-0.04}^{+0.04}$ & 0.63 $_{-0.04}^{+0.07}$ & 0.62 $_{-0.01}^{+0.02}$ & 0.62 $_{-0.04}^{+0.03}$ \\
       ($\rm RP_{hot}$ comp.) & $kT_{\rm init}$ (keV) & 5 (fixed) & 5 (fixed) & 5 (fixed) & 5 (fixed) & 5 (fixed) & 5 (fixed) & 5 (fixed) & 5 (fixed) & 5 (fixed) & 5 (fixed) \\
       & $Z_{\rm all}$ (solar) & = $\rm RP_{cold}$ & = $\rm RP_{cold}$ & = $\rm RP_{cold}$ & = $\rm RP_{cold}$ & = $\rm RP_{cold}$ & = $\rm RP_{cold}$ & = $\rm RP_{cold}$ & = $\rm RP_{cold}$ & = $\rm RP_{cold}$ & = $\rm RP_{cold}$ \\
       & $n_{e}t$ (10$^{11}$ s~cm$^{-3}$) & = $\rm RP_{cold}$ & = $\rm RP_{cold}$ & = $\rm RP_{cold}$ & = $\rm RP_{cold}$ & = $\rm RP_{cold}$ & = $\rm RP_{cold}$ & = $\rm RP_{cold}$ & = $\rm RP_{cold}$ & = $\rm RP_{cold}$ & = $\rm RP_{cold}$ \\
       & $VEM$ ($10^{57}~\rm cm^{-3}$)$^{\dagger}$ & 0.5 $_{-0.2}^{+0.3}$ & 0.1 $_{-0.1}^{+0.1}$ & 1.2 $_{-1.0}^{+1.1}$ & 0.5 $_{-0.1}^{+0.4}$ & 0.9 $_{-0.1}^{+0.2}$ & 0.3 $_{-0.1}^{+0.1}$ & 0.7 $_{-0.1}^{+0.2}$ & 0.3 $_{-0.1}^{+0.1}$ & 0.8 $_{-0.1}^{+0.1}$ & 0.3 $_{-0.1}^{+0.1}$ \\
      Gaussian & Centroid (keV) & 1.23 (fixed) & 1.23 (fixed) & 1.23 (fixed) & 1.23 (fixed) & 1.23 (fixed) & 1.23 (fixed) & 1.23 (fixed) & 1.23 (fixed) & 1.23 (fixed) & 1.23 (fixed) \\
       & Normalization$^{\ddagger}$ & 28 $_{-8}^{+8}$ & 5 $_{-5}^{+4}$ & 12 $_{-7}^{+7}$ & 15 $_{-4}^{+8}$ & 9 $_{-5}^{+5}$ & 11 $_{-8}^{+7}$ & 4 $_{-4}^{+4}$ & 4 $_{-2}^{+3}$ & 5 $_{-1}^{+1}$ & 6 $_{-2}^{+2}$ \\
      Power law & Photon index & ------------ & ------------ & ------------ & ------------ & ------------ & 1.89 (fixed) & 1.89 (fixed) & ------------ & ------------ & 2.22 (fixed) \\
       & Normalization$^{\ast}$ & ------------ & ------------ & ------------ & ------------ & ------------ & 45 $_{-9}^{+9}$ & 97 $_{-7}^{+7}$ & ------------ & ------------ & 18 $_{-5}^{+4}$ \\
      \hline
      & $\chi^{2}_{\nu}$ ($\nu$) &  1.18 (365) & 1.35 (250) & 1.25 (477) & 1.22 (365) & 1.53 (477) & 1.13 (364) & 1.36 (476) & 1.19 (273) & 1.27 (257) & 1.18 (256) \\
\enddata
\tablecomments{
$^{\dagger}$Volume emission measure at the distance of 1.5 kpc: $VEM = \int n_e n_{\rm H} dV$ , where $n_e$, $n_{\rm H}$, and $V$ are the electron and hydrogen densities, and the emitting volume, respectively.\\
$^{\ddagger}$The unit is photons s$^{-1}$ cm$^{-2}$ sr$^{-1}$.\\
$^{\ast}$The unit is photons s$^{-1}$ cm$^{-2}$ keV$^{-1}$ sr$^{-1}$ at 1~keV.\\
}
\end{deluxetable*}
\end{longrotatetable}

%

\end{document}